\documentclass[fleqn,usenatbib]{mnras}

\usepackage{newtxtext,newtxmath}
\usepackage[T1]{fontenc}
\usepackage{ae,aecompl}
\usepackage{mathtools}
\usepackage{subfigure}
\usepackage[labelfont=bf]{caption}

\usepackage{graphicx}	% Including figure files
\usepackage{amsmath}	% Advanced maths commands
\newcommand{\nn}{\nonumber}
\newcommand{\beq}{\begin{equation}}
\newcommand{\eeq}{\end{equation}}
\def\la{\buildrel < \over {_{\sim}}}
\def\ga{\buildrel > \over {_{\sim}}}

\title[Electron escape from SNRs]{Cosmic-ray electrons released by supernova remnants}
\author[Morlino \& Celli]{
	G.~Morlino$^{1}$ \thanks{giovanni.morlino@inaf.it}
	and S.~Celli $^{2,3}$  \thanks{silvia.celli@roma1.infn.it}
\\
    $^{1}$INAF, Osservatorio Astrofisico di Arcetri, L.go E.~Fermi 5, Firenze, Italy  \\
	$^{2}$Dipartimento di Fisica dell'Universit\`a La Sapienza, P.~le Aldo Moro 2, 00185 Rome, Italy \\
	$^{3}$Istituto Nazionale di Fisica Nucleare, Sezione di Roma, P.~le Aldo Moro 2, 00185, Rome, Italy \\
}

\begin{document}

\date{Accepted -----. Received -----}

%\pagerange{\pageref{firstpage}--\pageref{lastpage}} \pubyear{2011}

\maketitle

\label{firstpage}

\begin{abstract}
The process that allows cosmic rays to escape from their sources and be released into the Galaxy is still largely unknown. The comparison between cosmic-ray electron and proton spectra measured at Earth suggests that electrons are released with a spectrum steeper than protons by $\Delta s_{\rm ep} \sim 0.3$ for energies above $\sim 10$ GeV and by $\Delta s_{\rm ep} \sim 1.2$ above $\sim 1$~TeV. 
Assuming that both species are accelerated at supernova remnant shocks, we here explore two possible scenarios that can in principle justify steeper electron spectra: {\it i}) energy losses due to synchrotron radiation in an amplified magnetic field, and {\it ii}) time dependent acceleration efficiency.
We account for magnetic field amplification produced by either cosmic-ray induced instabilities or by magneto-hydrodynamics instabilities my means of a parametric description.
We show that both mechanisms are required to explain the electron spectrum. In particular synchrotron losses can only produce a significant electron steepening above $\sim 1$~TeV,
while a time dependent acceleration can explain the spectrum at lower energies if the electron injection into diffusive shock acceleration is inversely proportional to the shock speed. We discuss observational and theoretical evidences supporting such a behaviour. 
Furthermore, we predict two additional spectral features: a spectral break below $\sim$ few GeV (as required by existing observations) due to the acceleration efficiency drop during the adiabatic phase, and a spectral hardening above $\sim 20$~TeV (where no data are available yet) resulting from electrons escaping from the shock precursor. 
\end{abstract}


%%% SECTION %%%
\section{Introduction}  
\label{sec:intro}
The final spectrum of cosmic rays (CRs) as detected at Earth is determined by three different processes:  acceleration, escape from the sources and propagation through the Galaxy. Among the three, the escape process is the less understood one, partially because its comprehension relies on details of the acceleration process and magnetic field evolution, while being at the same time hard to constrain experimentally. In the context of the so-called SNR paradigm for the origin of CRs, which considers the bulk of Galactic CRs accelerated at supernova remnant (SNR) shocks, the spectral shape of electrons is usually assumed to be the same as protons, at least up the electron maximum energy, which is expected to be smaller than the proton one, because of the energy losses suffered during the acceleration process.

However, the spectrum of CR electrons (CRe) detected at Earth is remarkably different from the proton (CRp) one. The former, in fact, follows a power law in energy with a slope $\propto E^{-3.1}$ from $\sim 10$~GeV up to $\sim 1$~TeV (to be compared with the proton spectrum which is rather $\propto E^{-2.7}$). Above 1 TeV, different instruments have shown that a spectral steepening occurs in the energy distribution of electrons (CALET \citep{CALET:2018}, H.E.S.S. \citep{HESS_CRe:2008,HESS_CRe:2009,HESS_CRe:2017} and DAMPE \citep{DAMPE:2017}), that becomes compatible with a power law $\propto E^{-3.9}$ at least up to $\sim 20$ TeV, which is the highest energy at which electrons have been detected by H.E.S.S..

In a scenario where CRp and CRe are produced by the same sources, the different slopes below 1 TeV have been usually attributed to the energy losses suffered by electrons during the propagation through the Galaxy.
However, it is straightforward to see that these losses are not sufficient to explain the spectral deviation among them, if one correctly accounts for the different residence times that protons and electrons spend in the Galaxy, as we are going to show. In fact, defining $Q$ as the CR spectrum released by sources, the observed spectrum in the Galactic disk is $N \propto Q \,\tau/l$ where $l$ and $\tau$ are the propagation length and the propagation timescale, respectively. For protons $l$ is equal to the magnetic halo size $H$, and $\tau=H^2/D(E)$, $D$ being the diffusion coefficient. The propagation length for electrons is, instead, $l=\min[H,\sqrt{2 D \tau_{\rm loss}}]$ and, given the estimated halo thickness $H\gtrsim 5$\,kpc \citep{Evoli(All_CR):2019,  Evoli(Beryllium):2020, Weinrich+2020}, it is always dominated by losses due to inverse Compton (IC) scattering onto the Galactic photon background, at least for electron energies $\gtrsim 10$\,GeV \citep{Moskalenko-Strong:1998, Delahaye+2010, Evoli(electrons):2021}. Hence we have $N_e \propto Q_e \sqrt{\tau_{\rm loss}/D}$.
From AMS-02 measurements the proton-to-electron ratio is $N_p/N_e \sim E^{0.4}$, while the diffusion coefficient obtained from the combined fit of primary and secondary CR spectra is $D \propto E^{0.54}$ \citep{Evoli(All_CR):2019}. Finally, the energy loss timescale above $\sim 10$\,GeV can be approximated by a power-law $\tau_{\rm loss} \propto E^{-0.77}$ \citep{Evoli(electrons):2021}. As a consequence the ratio among injected spectra at the source is $Q_e/Q_p \propto N_e/N_p (D \tau_{\rm loss})^{-1/2} \propto E^{-\Delta s_{\rm ep}}$ with $\Delta s_{\rm ep} =0.28$.

Numerical solutions of the electron transport equation are in agreement with the above estimate. For example, \cite{diBernardo+2011} predict an electron spectrum injected by sources to be as steep as $\sim E^{-2.65}$ under the assumption of uniform source distribution.
In principle, non uniform source distribution can result into a less steep spectrum \citep{Gaggero+2013}, because of the fact that the Sun is located in a source under-dense region. This turns into a larger average distance travelled by electrons, hence stronger losses, between the bulk of sources in the arms and the observer.
However, more recent detailed calculations, including SNR locations following the spiral structure of Galactic arms and the contribution to CR electrons from pulsar wind nebulae (PWN) \citep{Evoli(electrons):2021, DiMauro+(electrons):2020}, are in good agreement with our estimate, indicating an electron spectrum steeper than the proton one by $\Delta s_{\rm ep} \simeq 0.3$. Possible corrections due to the presence of local closeby SNRs does not change significantly this conclusion \citep{Manconi+(electrons):2019}.

The most straightforward explanation for a steeper electron injection spectrum resides in the effect of energy losses that electrons suffer inside the sources, before being released into the interstellar medium (ISM). Such a framework has been investigated by \cite{Diesing-Caprioli2019}, \cite{Brose+2020} and \cite{Cristofari+2021}, who accounted for the synchrotron losses due to magnetic field amplification (MFA) through CR-streaming instability (SI). \cite{Cristofari+2021} concluded that SI is not sufficient to steepen the electron spectrum by the observed amount below $\lesssim 1$\,TeV. Such a conclusion might change only if additional MFA would be at work during the later stage of SNR evolution, when the bulk of low energy CR are produced. A result similar to \cite{Cristofari+2021} has been obtained by \cite{Brose+2020}, where a full numerical treatment has been used to get proton and electron spectra.
In turn, \cite{Diesing-Caprioli2019} reached a different conclusion: by adopting a different saturation for the SI, they obtain a large magnetic field even for low shock speed.
Such a saturation recipe is indeed suggested by particle-in-cell (PIC) simulations \citep{Caprioli-Spitkovsky:2014b}, but it results into maximum energies larger than what is estimated by gamma-ray observations of evolved SNRs.

It is worth noting that the saturation level of Bell instability is still a matter of debate \cite[see][for a detailed discussion]{Cristofari+2021}. In addition, other MFA mechanisms might be present beyond the SI. In particular magneto-hydrodynamics (MHD) instabilities are expected to arise when the shock propagates through a non uniform medium, giving rise to magnetic amplification downstream of the shock. The net effect of the development of such a magnetic field would be to increase the electron losses before their escape from the SNR, while not affecting the maximum energy reached at the shock.

The origin of the spectral break of CRe above $\sim 1$ TeV is also uncertain. A possible explanation resides into the energy losses experienced while still being located inside the sources. Alternatively, it could reflect the maximum energy at which electrons are accelerated \citep{Ohira_el:2012} or instead it could be due to the contribution of PWNe \citep{DiMauro+(electrons):2020, Evoli(electrons):2021}. It is even possible that it represents the sign of some different physical phenomena, e.g. the transition between a regime where a large number of sources contribute to the spectrum, to a regime where only a few, the closest ones to the Earth, are able to contribute. 
To this respect, \citet{Recchia(el)2019} and \citet{Fornieri+2020} have shown that a single local fading accelerator could be responsible for the CRe spectrum above $\gtrsim 1$\,TeV.

In principle, CRe and CRp could be produced by different sources able to accelerate them with different spectral index. However, it is unclear which kind of source can preferentially accelerate electrons given the observational constraints available. Pulsars are well known electron factories, but they also produce positrons with the same spectrum, hence they are excluded. An interesting possibility is provided by stellar winds, especially those in massive stellar clusters which have been recognized as gamma-ray sources \citep{Abramowski_Wd1:2012, Yang+2018, Aharonian+2019NatAs}. However, current interpretations of such emission tends to favour a hadronic origin, leaving the leptonic contribution unconstrained. Further work is needed to fully explore this scenario.

In this paper we assume that protons and electrons are both accelerated by SNRs by means of {\it diffusive shock acceleration} (DSA). We restrict our calculations to SNRs evolving into uniform circumstellar medium, postponing the case of remnants expanding in more complex environments to a future work.
We compute the proton and electron spectra released by an SNR by adopting a parametric description for the magnetic amplification mechanism, able to account for both CR-self generated (CR-SG) and turbulent amplification, such as to assess under which conditions steep electron spectra can be produced. Additionally, we explore a different mechanism that can produce steeper spectra, namely a time-dependent acceleration efficiency, and discuss evidences pointing towards an electron acceleration efficiency that is inversely proportional to the shock speed.

In \citet{Celli-escape1} (from now on Paper I) we calculated the CR proton spectrum produced by an SNR using a full analytical treatment, by solving the transport equation under the assumption that the SNR evolves according to the Sedov-Taylor solution. This was possible because protons only suffer adiabatic losses during the remnant expansion. In the case of electrons, the same approach cannot be used because of radiative losses, hence here we develop a more general technique which allows to calculate particle spectra along with the dynamical evolution of the remnant.
The paper is organized as follows. In \S~\ref{sec:evolution} we describe the temporal evolution of the SNR shock position and speed, which are essential ingredients to correctly model the behavior of shock-accelerated particles. These particles in fact satisfy the transport equation, however different conditions will apply to protons and electrons. The former are discussed in \S~\ref{sec:protons}, where we follow the methods introduced in Paper I for the temporal evolution of the maximum momentum of protons that the shock can confine. However, we here provide an improvement to that description, by extending its application to young remnants. Furthermore, we now include radiative losses of particles in both the self-amplified magnetic field of protons and the possible MHD turbulence developed downstream of the shock, as we detail in \S~\ref{sec:mag-field}. These losses are fundamental to correctly describe the evolution of electrons, which we provide in \S~\ref{sec:electrons}. We discuss the main differences among protons and electrons in \S~\ref{sec:results}, both in terms of maximum momentum achieved and injected spectrum inside the Galaxy, exploring few different scenarios that result in different spectral shape of the two particle populations. Finally, we conclude in \S~\ref{sec:conc}.

%%% SECTION %%%
\section{Supernova remnant evolution}    
\label{sec:evolution}
With respect to \cite{Celli-escape1}, where we only dealt with middle-aged SNRs, justifying the use of pure Sedov-Taylor solution, here we are going to adopt a fully numerical approach, hence we can accurately describe the dynamics of the SNR evolution through its transition from the ejecta-dominated (ED) to the Sedov-Taylor phase (ST) phase, following the parametrizations provided by \cite{tmk99}. We will deal only with the case of a remnant expanding into a uniform medium with mass density $\rho_0=n_0 m_{\rm p}$ ($m_{\rm p}$ being the proton mass and $n_0$ being the upstream proton numerical density). The time that marks the transition between these two phases is the so-called Sedov time, namely
\begin{equation} 
%t_\textrm{Sed} \simeq 1.6 \times 10^3 \, \textrm{yr} \left(\frac{E_\textrm{SN}}{10^{51} \, \textrm{erg}} \right)^{-\frac{1}{2}} \left(\frac{M_\textrm{ej}}{10 \, M\odot} \right)^{\frac{5}{6}}  \left(\frac{n_0}{\textrm{cm}^{-3}} \right)^{-\frac{1}{3}}
t_\textrm{Sed} \simeq 506 \, \textrm{yr} \left(\frac{E_\textrm{SN}}{10^{51} \, \textrm{erg}} \right)^{-\frac{1}{2}} 
   		\left(\frac{M_\textrm{ej}}{1 \, M\odot} \right)^{\frac{5}{6}} 
		\left(\frac{n_0}{0.1~\textrm{cm}^{-3}} \right)^{-\frac{1}{3}} \, ,
\end{equation}
where $M_\textrm{ej}$ is the mass ejected by the supernova (SN) explosion, and $E_{\rm SN}$ is the kinetic energy released at the SN. Note that the characteristic values adopted in this estimate refer to type Ia SN explosions, typically expanding into uniform density media. In such circumstances, during the ED stage, the shock speed is almost constant with time, while it significantly decreases with time after the SNR enters the ST stage. Introducing some characteristic scales of the problem, as a radius $R_{\rm ch}=M^{1/3}_\textrm{ej} \rho^{-1/3}_0$, a time $t_{\rm ch}=E^{-1/2}_\textrm{SN} M^{5/6}_\textrm{ej} \rho^{-1/3}_0$, and a speed $u_{\rm ch}=R_{\rm ch}/t_{\rm ch}$, the temporal evolution of the shock radius can be described through
\begin{equation}
\label{eq:ST_R}
\frac{R_{\rm sh}(t)}{R_{\rm ch}} = 
\begin{dcases}
2.01 \left(\frac{t}{t_{\rm ch}}\right) \left[1+1.72 \left(\frac{t}{t_{\rm ch}} \right)^{3/2} \right]^{-2/3} \qquad t<t_{\rm Sed} \\
\left[1.42 \left(\frac{t}{t_{\rm ch}} \right)-0.254 \right]^{2/5} \qquad \qquad \qquad t \geq t_{\rm Sed} \, ,
\end{dcases}
\end{equation}
while the shock speed by
\begin{equation}
\label{eq:ST_u}
\frac{u_{\rm sh}(t)}{u_{\rm ch}} = 
\begin{dcases}
2.01 \left[1+1.72 \left(\frac{t}{t_{\rm ch}} \right)^{3/2} \right]^{-5/3} \qquad \quad t<t_{\rm Sed} \\
0.569\left[1.42 \left(\frac{t}{t_{\rm ch}} \right)-0.254 \right]^{-3/5} \qquad t \geq t_{\rm Sed} \, ,
\end{dcases}
\end{equation}
both holding for a remnant expanding into a uniform density profile of the circumstellar medium and a constant structure function for the ejecta velocity \citep{tmk99}.

The calculation of adiabatic losses requires the knowledge of the internal structure of the SNR. Here we adopt the linear velocity approximation introduced by \citet{Ostriker-McKee1988}, and also adopted by \citet{Ptuskin+2005} as well as in Paper I, in which the plasma velocity profile for $r \leq R_{\rm sh}$ is given by
\begin{equation}
\label{eq:ush}
    u(t,r) = \left( 1 - \frac{1}{\sigma}\right) \frac{u_{\rm sh}(t)}{R_{\rm sh}(t)} r \,,
\end{equation}
$\sigma$ being the compression ratio at the shock.

Finally, the SNR transits to the pressure-driven snowplough phase when radiative losses become important. Following the calculation of \cite{Cioffi+1988} we estimate this transition time as
\begin{equation}
\label{eq:t_SP}
%t_{\rm SP} = 1.33 \times 10^4 \, E_{\rm SN,51}^{3/14} n_{0,0}^{-4/7} \zeta_{m}^{-5/14} \; {\rm yr} \,,
t_{\rm SP} = 4.95 \times 10^4 \, \left( \frac{E_{\rm SN}}{10^{51}~{\rm erg}} \right)^{\frac{3}{14}} \left(\frac{n_0}{0.1~\textrm{cm}^{-3}} \right)^{-\frac{4}{7}} \left( \frac{\zeta_{m}}{1} \right)^{-\frac{5}{14}} \; {\rm yr} \,,
\end{equation}
where $\zeta_m$ is a dimensionless correction factor to account for metallicity variation with respect to Solar abundances (corresponding to $\zeta_m = 1$).  In the following calculations we will assume that particle acceleration stops at $t_{\rm SP}$: in fact, as we showed in Paper I, efficient acceleration during the snowplough phase would result into a hard CR proton spectrum at $E \lesssim 10$~GeV (due to the different scaling of shock velocity with time), which however is in contrast with CR observations.
In addition, radio emission of shell-type SNRs suggests that electron acceleration stops when the evolution enters this stage \citep{Bandiera-Petruk:2010}.
However, from a theoretical point of view, the reason why acceleration should stop is not clear, given that the Mach number is usually still $\gtrsim 10$. Possible explanations could be related to the fragmentation of the dense shell behind the shock \citep{Blondin+1998}, which would break the shock thus allowing particles to escape, or to the presence of a large fraction of neutral hydrogen, which would result into damping of magnetic turbulence and subsequent drop of the acceleration efficiency.
Another possibility is related to the fact that the SNR evolution during the radiative phase may be modified by the CR pressure  \citep{Diesing-Caprioli2018} in such a way that acceleration may proceed without producing spectral features. However, all these hypotheses remains to be proven.

%%% SECTION %%%
\section{Proton spectrum}
\label{sec:protons}
In this Section, we summarize all the ingredients necessary for the description of the particle diffusion model that we developed in Paper I. For further details the reader is referred to such paper. 
We assume spherical symmetry both inside and outside the SNR, such that the particle distribution function $f(t,r,p)$ is described by the time-dependent transport equation in spherical coordinates
\begin{equation} 
\label{eq:transport}
 \frac{\partial{f}}{\partial{t}} + u \frac{\partial{f}}{\partial{r}} = \frac{1}{r^2} \frac{\partial}{\partial r} \left[ r^2 D \frac{\partial{f}}{\partial{r}}\right]
 + \frac{1}{r^2} \frac{\partial (r^2 u)}{\partial r}  \frac{p}{3} \frac{\partial f}{\partial p} \, ,
\end{equation}
where $u$ is the  plasma velocity and $D$ the particle diffusion coefficient. In solving Eq.~(\ref{eq:transport}) we distinguish between \emph{confined} and \emph{non-confined} particles. The former population encloses all particles whose momentum $p$ is lower than the maximum momentum of particles accelerated at the shock at each time $t$, i.e. $p_{\max,0}(t)$. These particles are attached to the expanding plasma, and as such they undergo advection and adiabatic losses. The \emph{non-confined} population encloses, instead, all remaining particles, namely those that have escaped the shock, and can freely diffuse away from the source. Concerning the maximum energy at the shock, following Paper I, we will parametrize its dependence over time in the form of
\begin{equation} 
\label{eq:pmax0}
 p_{\max,0}(t) =
  \begin{cases} 
   p_\textrm{M} \left( t/t_{\rm Sed} \right)     & \text{if } t \leqslant  t_{\rm Sed} 	\\
   p_\textrm{M} \left( t/t_{\rm Sed} \right)^{-\delta}     & \text{if } t > t_{\rm Sed} \,,
  \end{cases}
\end{equation}
where $p_\textrm{M}$ represents the maximum momentum, achieved at $t=t_{\rm Sed}$, while the slope $\delta$ is a free parameter. We note that the value of $\delta$ estimated from multi-wavelength fitting of our model to two  middle-aged SNRs, namely Cygnus Loop \citep{Loru+2021} and Gamma-Cygni \citep{MAGIC-GammaCigni2020}, ranges between 2 and 3, hence this range will be assumed for reference.
By inverting Eq.~\eqref{eq:pmax0} we can also define the so-called escape time $t_{\rm esc}(p)$, namely the time at which particles with momentum $p$ escape the SNR:
\begin{equation} 
\label{eq:tesc}
t_{\rm esc} (p) = t_{\rm Sed} (p/p_{\rm M})^{-1/\delta}
\end{equation}
and the corresponding escape radius $R_{\rm esc}(p)\equiv R_{\rm sh}(t_{\rm esc}(p))$.

The method we adopted to estimate the escape time relies on the idea that if particles are not confined at the shock, they cannot be confined inside the SNR either because the magnetic field inside the SNR is always smaller than the one at the shock due to its adiabatic expansion and damping (see \S~\ref{sec:mag-field}). However, a more rigorous method for estimating the escape time should also account for the diffusion coefficient self-generated by the confined particles while these are diffusing away. Such a calculation was performed by e.g. \cite{Nava+2016, Nava+2019}, who also accounted for several mechanisms possibly responsible for the damping of magnetic turbulence. Interestingly, their results are in agreement with our recipe assuming $\delta \simeq 2 \div 3$ \cite[see also][figure 4 and related discussion]{Recchia+2021}.

The distribution function of CR accelerated at the shock, $f_0(p,t)$, is determined by DSA and it is described by a power-law in momentum suppressed by an exponential cut-off:
\begin{equation} 
\label{eq:f_0}
 f_0(p,t) = \frac{3 \, \xi_{\rm CR,p} \, u^2_{\rm sh}(t) \rho_0}{4 \pi \, c (m_{\rm p} c)^4  \Lambda(p_{\max,0}(t))} 
 		\left( \frac{p}{m_{\rm p} c} \right)^{-\alpha}  \exp \left[ -\frac{p}{p_{\max,0}(t)} \right] \, ,
\end{equation}
$c$ being the speed of light in vacuum,while $\alpha$ is the spectral slope, related to the acceleration process ($\alpha=4$ is expected in the test-particle regime of DSA). The proton acceleration efficiency $\xi_{\rm CR,p}$ is assumed constant in time, while the normalization factor $\Lambda$ is calculated imposing that the CR pressure at the shock satisfies $P_{\rm CR} = \xi_{\rm CR,p} \rho_0 u_{\rm sh}^2$.
The subsequent evolution of proton spectrum is calculated as for electrons, see \S~\ref{sec:el_conf} and \ref{sec:el_esc}, but neglecting radiative losses.

%%% SECTION %%%
\section{Magnetic field evolution}	
\label{sec:mag-field}
The presence on non-thermal electrons in SNRs is revealed by numerous observations of radiation spanning from radio to X rays, proving the shock capabilities to accelerate electrons all the way from GeV to TeV \citep{vink2012}. Such emissions clearly constitute an energy loss process for electrons. In order to evaluate its impact as a function of time, we need to estimate the magnetic field strength at the shock, and its evolution in the remnant interior during its expansion. The value of the magnetic field at the shock is the result of both amplification and compression at the shock of the circumstellar magnetic field. Since we here aim at describing a picture that might be applied to a broad variety of SNRs, in different evolutionary stages, we do not attempt to explicitly describe any particular amplification mechanism, rather we account for MFA parametrically, by distinguishing between two different categories: amplification induced by CR-streaming instabilities, also known as self amplification (\S~\ref{sec:mag-field_SG}), and MHD instabilities produced by the plasma motion (\S~\ref{sec:mag-field_MHD}). 
Both mechanisms affect electron losses, but here these are treated separately because of their different dependence on the shock speed.

%%% SUB-SECTION %%%
\subsection{CR-generated magnetic field}
\label{sec:mag-field_SG}
The magnetic field self-generated by CRs is connected to the maximum energy of protons, as given by Eq.~\eqref{eq:pmax0}.
Assuming that $p_{\max,0}(t)$ is determined by the age-limited condition $t_{\rm acc} = t_{\rm SNR}$, we can derive the magnetic field using the acceleration time $t_{\rm acc} \simeq 8 D_1(p)/u_{\rm sh}^2$ where the upstream diffusion coefficient is $D_1(p) = D_{B}/\mathcal{F}$, $D_B$ being the Bohm diffusion coefficient and $\mathcal{F}$ the magnetic logarithmic power spectrum. Note that we will use the subscript 1 (2) for the quantities calculated in the upstream (downstream). Because of Eq.~\eqref{eq:pmax0}, we get
\beq  
\label{eq:F}
\mathcal{F}(t) = \frac{8 \, p_{\rm M} c}{3 \, e B_0 \, c \, t_{\rm Sed}} 
\begin{dcases}
\left( \frac{u_{\rm sh}}{c} \right)^{-2} \qquad \qquad \qquad \quad t<t_{\rm Sed} \\
\left( \frac{u_{\rm sh}}{c} \right)^{-2} \left( \frac{t}{t_{\rm Sed}} \right)^{-\delta - 1} \qquad t \geq t_{\rm Sed} \, ,
\end{dcases}
%  \mathcal{F}(t) = \frac{8 \, p_{\rm M} c}{3 \, e B_0 \, c \, t_{\rm Sed}} \left( \frac{t}{t_{\rm Sed}} \right)^{-\delta - 1}
%$$  \left( \frac{u_{\rm sh}(t)}{c} \right)^{-2} \,.
\eeq
where $B_0$ is the upstream ordered magnetic field. E.g., during the ST phase $u_{\rm sh} \propto t^{-3/5}$, implying that $\mathcal{F} \propto t^{-\delta+1/5}$.
When the magnetic field is amplified beyond the linear regime ($\delta B \gtrsim B_0$), we consider the diffusion as Bohm-like in the amplified magnetic field. In other words, the function $\mathcal{F}$ behaves as $\mathcal{F} \sim (\delta B/B_0)^2$ for $\delta B \ll B_0$  and $\mathcal{F} \sim (\delta B/B_0)$ for $\delta B \gg B_0$ \cite[see, e.g.][]{Blasi_review:2013}. 
We assume an empirical formula reproducing these limits, namely $\mathcal{F}^{-1} = (B_0/ \delta B) + (B_0/ \delta B)^2$, which once inverted provides
\beq  
\label{eq:deltaB_F}
  \delta B_1(t) = \frac{B_0}{2} \left( \mathcal{F}(t) + \sqrt{4 \mathcal{F}(t) + \mathcal{F}^2(t)} \right) \,.
\eeq
The total magnetic field strength in the shock upstream is then $B_{1,\rm tot}(t) = \left(B_0^2 + \delta B_1^2(t) \right)^{1/2}$. Crossing the shock towards downstream, the magnetic field is further compressed by a factor $r_{\rm B}$, in both the oriented and the turbulent component: e.g. for a randomly oriented field, the average compression factor is $r_{\rm B}=\sqrt{11}$. As a result, the downstream total field at the shock position is equal to $B_{2,\rm tot}(t) = r_{\rm B} B_{1,\rm tot} (t)$.

In addition to field compression, the evolution of the downstream field is further affected by adiabatic losses and possible damping processes.
Several damping mechanisms have been proposed as effective in SNRs (see e.g. \citet{Ptuskin-Zirakashvili:2003}, and references therein), but here we focus only on non-linear damping, which is expected to be fairly efficient in hot plasmas, as outlined by \citet{volk1981} and \citet{mckenzie1982}.
We therefore assume that the magnetic field in the downstream is damped at a rate given by \cite[see Eqs.~(10)-(12) in][]{Ptuskin-Zirakashvili:2003}:
\beq
  \Gamma_{\rm nld} (k,t) = {(2 c_k)}^{-3/2} k v_{\rm A}(t) {\mathcal{F}(t)}^{1/2} \, ,
  \label{eq:Gamma_damp}
\eeq
$c_k=3.6$ being the Kolmogorov constant. Note that the Alfv\'en speed $v_{\rm A}=B_{2}(t)/\sqrt{4\pi n_{\rm i} m_{\rm i}}$ (for a ionized medium of numerical density $n_{\rm i}$ composed of ions with mass $m_{\rm i}$) depends on time because of the ordered component of the downstream magnetic field $B_2(t)$. Since the physical scale that dominates the cascade of turbulence resulting from the damping process is the largest, we will only consider a wavenumber $k$ comparable to the inverse of the Larmor radius of the highest energy protons, i.e. $k_{\rm res}=1/r_{\rm L}(p_{\rm max,0})$. 
Concerning the downstream magnetic field, the ordered component gets diluted with position within the shock radius and time as \citep{Reynolds:1998}
\beq
  B^2_2(r,t)=\frac{B^2_0}{3} \left[ \left(\frac{R_{\rm sh}(t)}{r} \right)^4 + 2 \sigma^2 L^6(t^\prime, t) \left(\frac{R_{\rm sh}(t)}{r} \right)^2 \right] 
  \label{eq:B2} \, ,
\eeq
holding in the assumption of an isotropic magnetic field. Note that $\sigma=u_1/u_2$ is the shock compression ratio ($\sigma =4$ for a strong shock), while $t^\prime(t,r)$ indicates the time when the plasma located at time $t$ in position $r$ was shocked. In order to determine this time, we follow an approach analogous to that of Paper I, where we could find an analytical solution for $t^\prime$ during the ST stage (see Eq.~(18) in Paper I). On the other hand, since we are here describing in details also the ED stage, we now switch towards a numerical solution, whose details are provided in Appendix~\ref{sec:appA}.
%Basically, the time when the plasma located at time $t$ in position $r$ was shocked follows from the assumption of the radial profile of the plasma velocity inside the remnant \citep{Ptuskin+2005}, as we give in Eq.~\eqref{eq:ush}.
%\beq
%\label{eq:vPtuskin}
%v(r,t)=\left(1-\frac{1}{\sigma} \right) \frac{r}{R_{\rm sh}} v_{\rm sh} 
%\eeq
%By integrating such equation between $R_{\rm sh}(t^\prime)$ and $r$ (respectively $t^\prime$ and $t$), and accounting for the temporal dependence of the shock radius and speed as given in Eqs.~\eqref{eq:ST_R}-\eqref{eq:ST_u}, we end up with an implicit equation in $t^\prime$. We refer to Appendix~\ref{sec:appA} for further details on the its derivation. 

The factor $L(t^\prime,t)$ in Eq.~\eqref{eq:B2} accounts for adiabatic losses that the magnetic field undergoes in the time interval $t-t^\prime$, and it is defined as \begin{equation}
\label{eq:L}
  L(t^\prime,t) =  \left[ \frac{\rho_2(t,r)}{\rho_{2}(t^\prime(t,r))} \right]^{1/3} \, ,
\end{equation}
where $\rho_{2}(t^\prime,r)$ is the density of the downstream (shoked) plasma element right at the time it was shocked, i.e. $t^\prime(t,r)$. Note that particles are subject to the same losses, as we are going to describe in \S~\ref{sec:electrons}. To compute the adiabatic factor, for which it holds that $L(t^\prime,t) \leq 1$, we make use of its dependence on the shock radius as
\begin{equation}
  L(t^\prime,t) = \left[ \frac{R_{\rm sh}(t^\prime)}{R_{\rm sh}(t)} \right]^{3/4} \, .
\end{equation}

The turbulent component of the magnetic field, $\delta B_2$, suffers both adiabatic losses and damping, hence its evolution can be described as:
\beq
  \delta B^2_2(t,r)=\frac{\delta B^2_1}{3} \left[ \left(\frac{R_{\rm sh}}{r} \right)^4 + 2 \sigma^2 L^6(t^\prime, t) \left(\frac{R_{\rm sh}}{r} \right)^2 \right] e^{- \left(t-t^\prime \right) \Gamma_{\rm nld} (t)} \,.
  \label{eq:dB2}
\eeq
In conclusion, the total magnetic field strength can simply be computed by summing in quadrature the ordered and the turbulent components, resulting into
\begin{equation}
\begin{split}
B_{2, {\rm tot}}(t,r) & =  \left[ \left(\frac{R_{\rm sh}(t)}{r} \right)^4 + 2 \sigma^2 L^6(t^\prime, t) \left(\frac{R_{\rm sh}(t)}{r} \right)^2 \right]^{1/2} \times \\
& \times \sqrt{ \frac{1}{3} \left[ B^2_0+ \delta B^2_1 e^{- \left(t-t^\prime \right) \Gamma_{\rm nld}}  \right] } \,.
\end{split}
\label{eq:Be_evol}
\end{equation}

%%% SUB-SECTION %%%
\subsection{Magnetic field from turbulent amplification}
\label{sec:mag-field_MHD}
Additional magnetic turbulence in the region downstream of the shock can be generated through the Richtmeier-Meshkov instability, namely a turbulent dynamo due to shock rippling when the SNR is expanding into a non homogeneous medium \citep{Giacalone-Jokipii:2007,Inoue:2012,Celli-clumps:2019}. In fact, in the presence of density inhomogeneities, vorticity may develop after the passage of the shock, giving rise to an enhancement  of  the  local  magnetic field with a strength which depends on the shock speed and upstream density contrast. 
%The  generated  turbulence then cascades to smaller spatial scales through a Kolmogorov-like process on timescales \silvianote{$\tau_{\rm cascade} = L / v_{\rm A}$. E.g. in the case of molecular clumps with $L \sim 0.1$~pc, embedded in low density cavities, we expect $\tau_{\rm cascade} \simeq 50$ yr} \cite[see, e.g.][]{}. 
%where $R_{\rm c}$ is the clump size and $v_{\rm A}$ is the Alfv\'en speed in the lower density medium 

Because such a turbulence only develops in the shock downstream, its intensity does not alter the value of the particle maximum momentum $p_{\max,0}$, which is instead connected with the self-generated turbulence upstream of the shock. However, the effect of MHD turbulence might be significant in terms of particle energy losses by synchrotron radiation.

Here we account for the possible presence of  turbulent amplification by using a simplified approach where the magnetic energy density immediately downstream of the shock is assumed to amount to a fixed fraction $\xi_{\rm B}$ of the shock kinetic energy, namely
\begin{equation}
    \frac{\delta B_{2,\rm tur}^2}{8\pi} = \xi_{\rm B} \frac{1}{2} \rho u_{\rm sh}^2 \,.
\end{equation}
At later times the evolution of this component is calculated following the same procedure described in \S~\ref{sec:mag-field}, hence we use Eq.~\eqref{eq:dB2} with the same damping rate but without compression (i.e. $\sigma=1$).

%%% SECTION %%%
\section{Electron spectrum} 	
\label{sec:electrons}

%%% SUBSECTION %%%
\subsection{Spectrum at the shock} 
\label{sec:el_sh}
The instantaneous electron spectrum at the shock, $f_{e,0}(p)$, is assumed proportional to the proton spectrum. Nonetheless, its cutoff is located at the maximum energy which is determined by the condition $t_{\rm acc} = \min[t_{\rm SNR}, \tau_{\rm loss}]$. In addition, the cutoff shape is different whether the maximum energy is limited by the source age $t_{\rm SNR}$ or by energy losses occurring on timescale $\tau_{\rm loss}$. In the former case the cutoff can be approximated by a pure exponential shape, just like the proton spectrum, i.e.
\begin{equation} 
\label{eq:f_e0_1}
 f_{\rm e,0}(p) = K_{\rm ep}\,f_{\rm p,0}(p) \, e^{-\left( \frac{p}{p_{\rm max,0}} \right)} \,, 
\end{equation}
where the cutoff momentum is defined in Eq.~\eqref{eq:pmax0}. The factor $K_{\rm ep}$ accounts for the different normalization between electrons and protons, which is likely related to the different mechanisms responsible for lepton and hadron injection. In the following, $K_{\rm ep}$ will be assumed constant and equal to unity, since we are not interested in the absolute relative normalization between CRe and CRp. Nonetheless, $K_{\rm ep}$ might possibly be a function of the shock speed, as we will discuss more in details in \S~\ref{sec:results_Kep}.

In the loss dominated case, a super-exponential cutoff is present \citep{Zirakashvili-Aharonian:2007,Blasi:2010,Yamazaki+2015}. In particular, when energy losses are proportional to $E^2$, like in the case of synchrotron and IC processes, the cutoff is $\propto \exp{[-(p/p_{\max,e})^2]}$. 
%, where the subscript L in the maximum momentum indicates the loss-dominated scenario
A good approximation to the spectrum of electrons is then provided by the expression worked out by \cite{Zirakashvili-Aharonian:2007}:
\begin{equation} 
\label{eq:f_e0_2}
f_{\rm e,0}(p) = K_{\rm ep} \, f_{\rm p,0}(p)  {\left[1+0.523 \left(p/p_{\rm max,e}\right)^{\frac{9}{4}}\right]}^2 \, e^{- \left(\frac{p}{p_{\rm max,e}} \right)^2} \, . \\
\end{equation}
Such a spectrum is expected to be a reasonable approximation of the true one \citep[see, e.g.,][]{Blasi:2010}, since the accelerated proton spectrum is considered to be a power law $\propto p^{-4}$ or slightly steeper.

We now estimate the electron maximum energy, as possibly limited by energy losses. The energy loss rate due to synchrotron plus IC scattering is
\begin{equation}
 \left( \frac{{\rm d} E}{{\rm d} t} \right)_{\rm syn+IC} = -\frac{\sigma_{\rm T} c}{6 \pi} \left( \frac{E}{m_{\rm e} c^2}\right)^2 \left( B^2 + B_{\rm eq}^2 \right) \, ,
\end{equation}
where $\sigma_{\rm T}$ is the Thompson cross section and $m_{\rm e}$ the mass of the electron, while $B_{\rm eq}^2 =  8\pi U_{\rm rad}$ is the equivalent magnetic field associated to the interstellar radiation field (ISRF) of energy density $U_{\rm rad}$. We assume a multi-component ISRF made of the cosmic microwave background (CMB), infrared (IR), optical (OPT), and ultraviolet (UV) components. The intensity of these radiation fields strongly depend on the region of the Galaxy where the source is located \cite[see, e.g.][]{Porter+:2017, Vernetto-Lipari:2016}, being generally enhanced towards the Galactic Center (GC). In the following, we will assume a source at a distance $d=4$~kpc from the GC, taking consequently the energy density of each component equal to $U_{\rm CMB} = 0.26\, \rm eV \,cm^{-3}$, $U_{\rm IR} = 0.02 \,\rm eV \,cm^{-3}$, $U_{\rm OPT} = 0.20 \,\rm eV \, cm^{-3}$, and $U_{\rm UV} = 0.43 \,\rm eV \, cm^{-3}$, or in other words an equivalent magnetic field of $B_{\rm eq}=4.8\, \mu$G.

\begin{figure}
\centering
\includegraphics[width=0.47\textwidth]{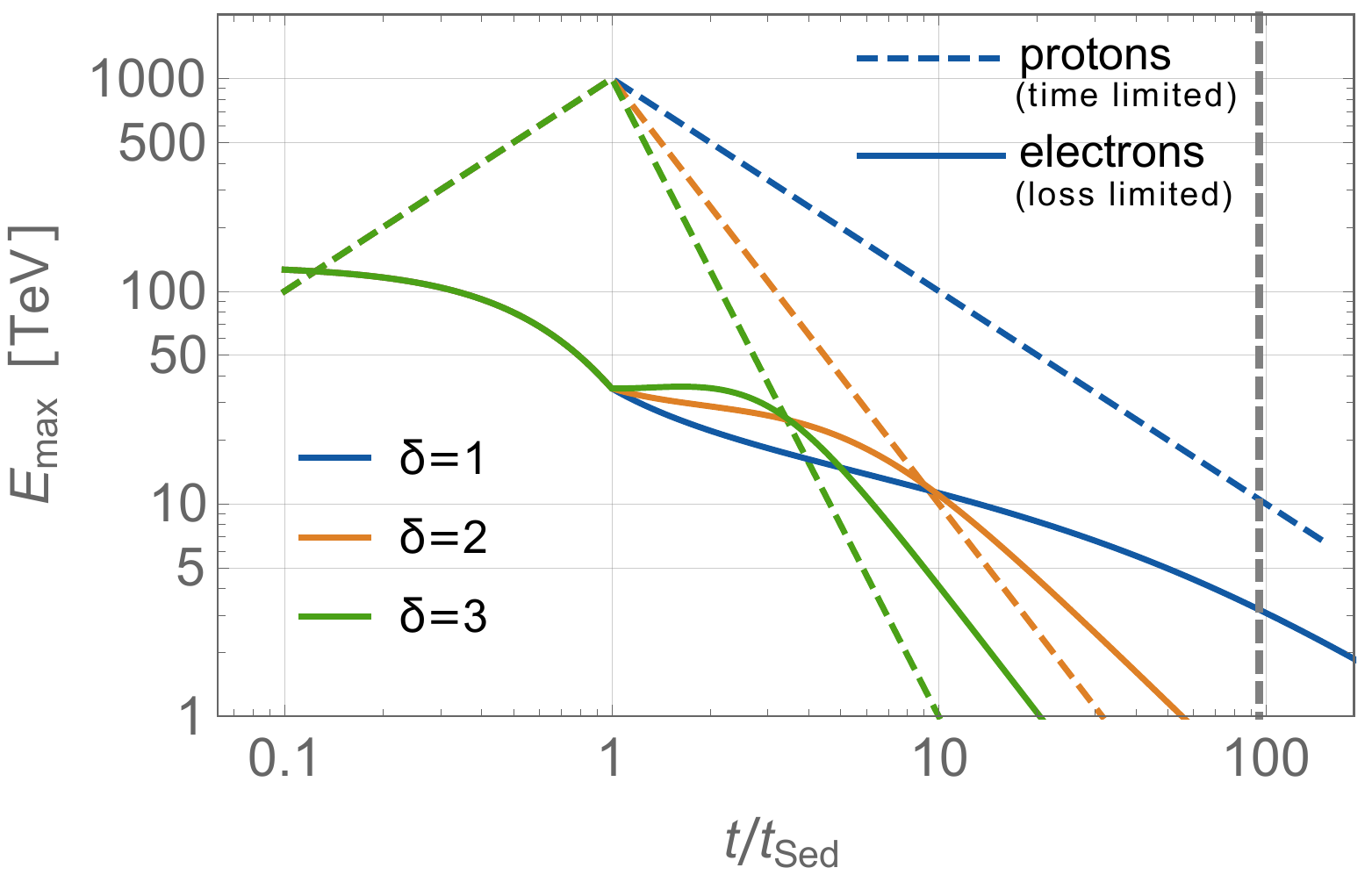}
\caption{Maximum energy of electrons at the shock as a function of time, determined by the loss limited condition (solid line) compared with the maximum energy of protons (dashed lines) for different values of $\delta$. The figure is obtained with parameters values of Table~\ref{tab:tab1} and for $p_{\rm M}= 1$ PeV/c. The vertical dashed-gray line show the beginning of the radiative phase.}
\label{fig:Emax_e} 
\end{figure}

The total loss timescale is calculated from the synchrotron + IC losses, averaged over the time spent upstream and downstream of the shock, namely, 
\beq
\label{eq:tau_loss_tot}
  \tau_{\rm loss} = \frac{t_{\rm res,1} + t_{\rm res,2}}{ t_{\rm res,1}/\tau_{\rm loss,1} + t_{\rm res,1}/\tau_{\rm loss,2}} \, ,
\eeq
the residence time being $t_{\rm res, i}=D_i/(c\,u_i)$. Imposing the condition $t_{\rm acc} = \tau_{\rm loss}$ we get the following expression for the maximum energy
\beq
\label{eq:Emax_e_loss}
  \frac{E_{\rm max,e}(t)}{m_e c^2}= \sqrt{ \frac{(\sigma-1) r_{\rm B}}{\sigma \left[ r_{\rm B} (1+\sigma_{\rm eq}^2) + \sigma(r_{\rm B}^2 + \sigma_{\rm eq}^2) \right]} \, \frac{6\pi e B_0 \mathcal{F}(t)}{\sigma_{\rm T} B_{1,\rm tot}^2(t)}} \; \frac{u_{\rm sh}(t)}{c} 
\eeq
where $\sigma=4$, $r_{\rm B}=\sqrt{11}$, and $\sigma_{\rm eq} = B_{\rm eq} / B_{1,\rm tot}$. It is worth noting that when the magnetic field is strongly amplified, the electron maximum energy is almost constant. In fact, for $\mathcal{F} \gg 1$, the IC losses can be neglected and the time dependence of the maximum energy is $E_{\rm max,e} \propto u_{\rm sh}(t) \mathcal{F}(t)^{-1/2} \propto t^{\delta/2 - 7/10}$. Hence, for small values of $\delta (\sim 1\div 2$) the time dependence is very mild, while for larger value ($\delta \gtrsim 3$) the time dependence becomes stronger but the loss limited condition only applies for a short time interval. As a result, for $t>t_{\rm Sed}$, we always have $E_{\rm max,e}$ smaller than $\sim 40$~TeV, as can be appreciated in Figure~\ref{fig:Emax_e}, which shows the electron maximum energy resulting from the energy loss condition, as compared with the proton maximum energy,  for the benchmark case summarized in Table~\ref{tab:tab1}, and several values of $\delta$. Among these, we also show the case $\delta=1$, which represents the expectation in the presence of MFA due to resonant streaming instability only (see Appendix~A in Paper I for a theoretical estimate of this parameter).
Note that for $t<t_{\rm Sed}$ the maximum energy of electrons lowers as the shock speed decreases differently from protons. 

\begin{figure*}
\centering
\subfigure[]{\includegraphics[width=0.47\textwidth]{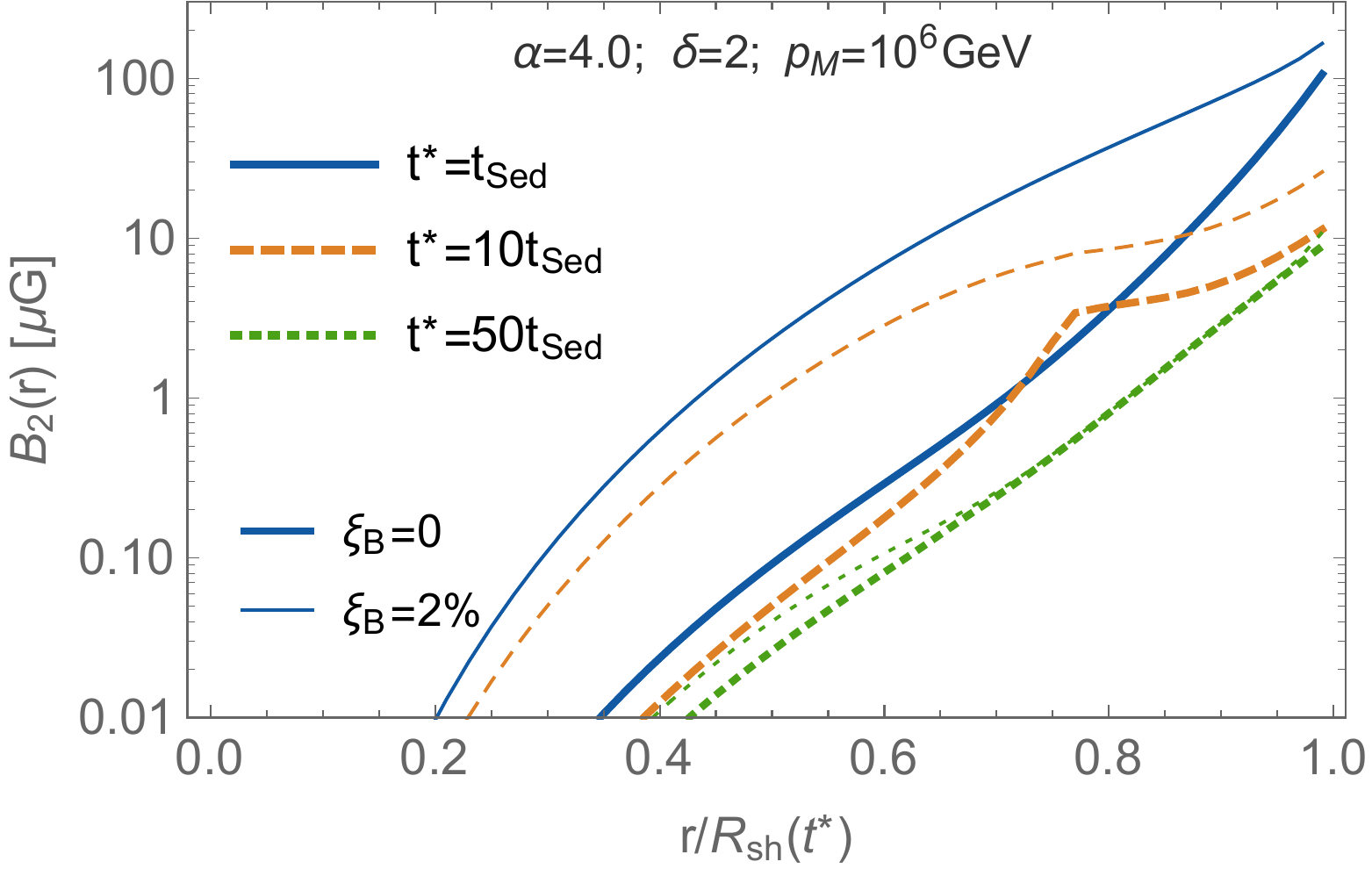}}
\subfigure[]{\includegraphics[width=0.47\textwidth]{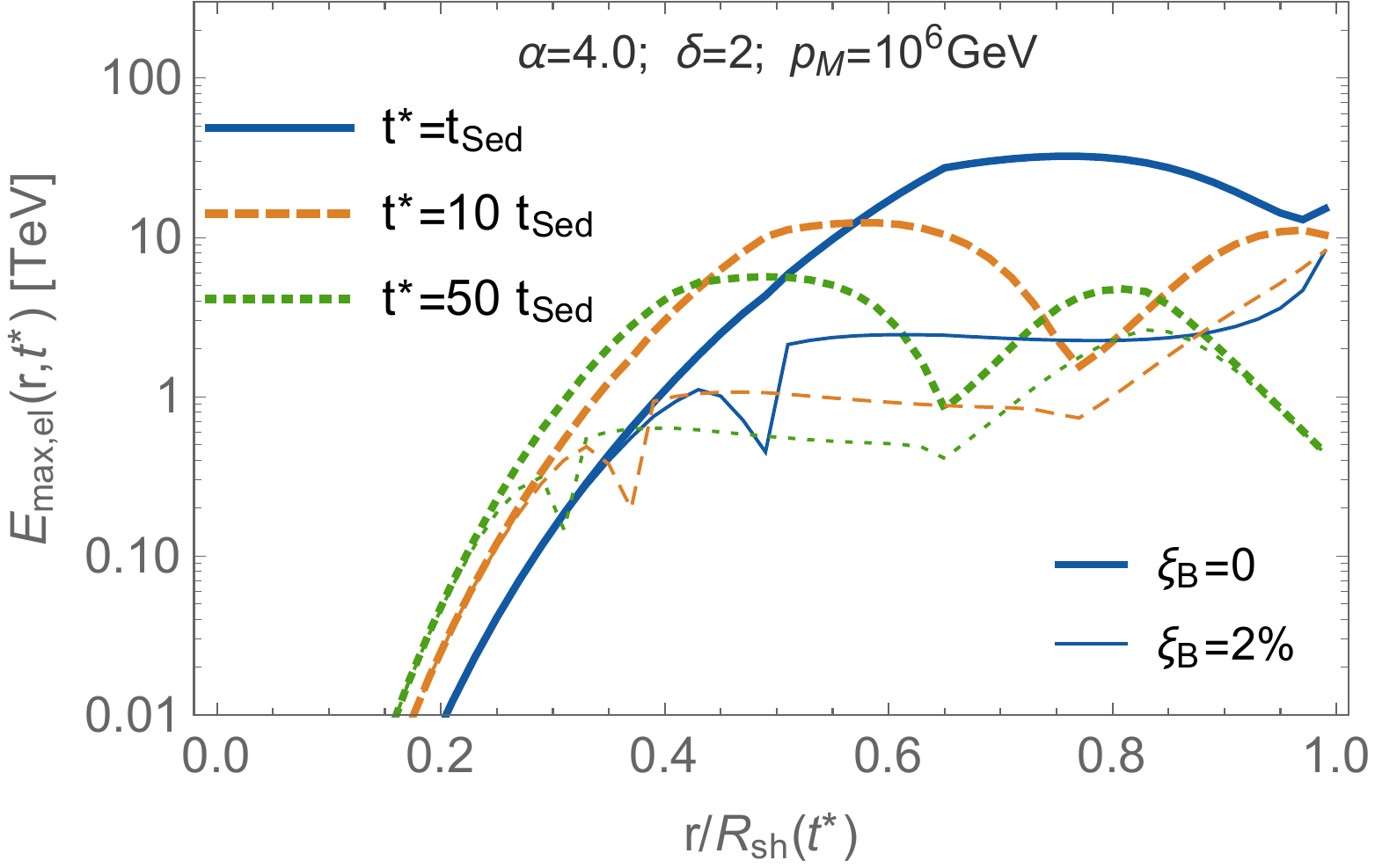}}
\caption{(a) Total magnetic field and (b) maximum energy of electrons in the SNR interior as a function of the $r/R_{\rm sh}(t^*)$ for three different times $t^*/t_{\rm Sed}=1$, 10 and 50. In both panels, thick lines refer to self-generated magnetic field only, while thin lines also account for MHD amplification with $\xi_{\rm B}=2\%$.}
\label{fig:EmaxBdown} 
\end{figure*}

For the same benchmark case of Table~\ref{tab:tab1}, we also show in Figure~\ref{fig:EmaxBdown} the radial profile of the maximum energy of electrons, as well as the downstream magnetic field, evaluated at several times. To evaluate the effect of turbulent magnetic field amplification, we also show the case with $\xi_{\rm B}=2\%$ (thin lines). 
Interestingly, for $\xi_{\rm B}=0$, the maximum energy is not a monotonic function of the radius, because of the combined effect between the time evolution of magnetic field and $p_{\max,e}(t)$:  as such, at a fixed age of the remnant, it is possible to find the highest energy particles both at the shock position as well as in other inner positions.
When the turbulent amplification is also included, the electron energy inside the SNR decreases more rapidly, with the exception of later times when the amplified magnetic field is small and losses are dominated by adiabatic expansion.

%%% SUBSECTION %%%
\subsection{Distribution of confined electrons} 
\label{sec:el_conf}
Once the electron spectrum at the shock is known, we can proceed to calculate the evolution of the particle distribution function in the remnant downstream considering the energy losses due to both adiabatic expansion and radiative processes. The equation describing the temporal evolution of the particle energy in such a  case reads as
\begin{equation}
 \frac{{\rm d} E}{{\rm d} t} = \left( \frac{{\rm d} E}{{\rm d} t} \right)_{\rm syn+IC} + \frac{E}{L}\frac{{\rm d}L}{{\rm d}t} \, ,
\end{equation}
where $L$ is the adiabatic loss function, given in Eq.~\eqref{eq:L}. Note that, as we did in the case of protons, we assume here that electrons are confined in their plasma elements and do not diffuse away from it. This is a good approximation if the typical diffusion length is much smaller than the SNR size, namely $\sqrt{D_{\rm in} \, t} \ll R_{\rm sh} (t)$ (see discussion in Section 2.4 of Paper I).

\begin{figure}
\centering
\includegraphics[width=0.47\textwidth]{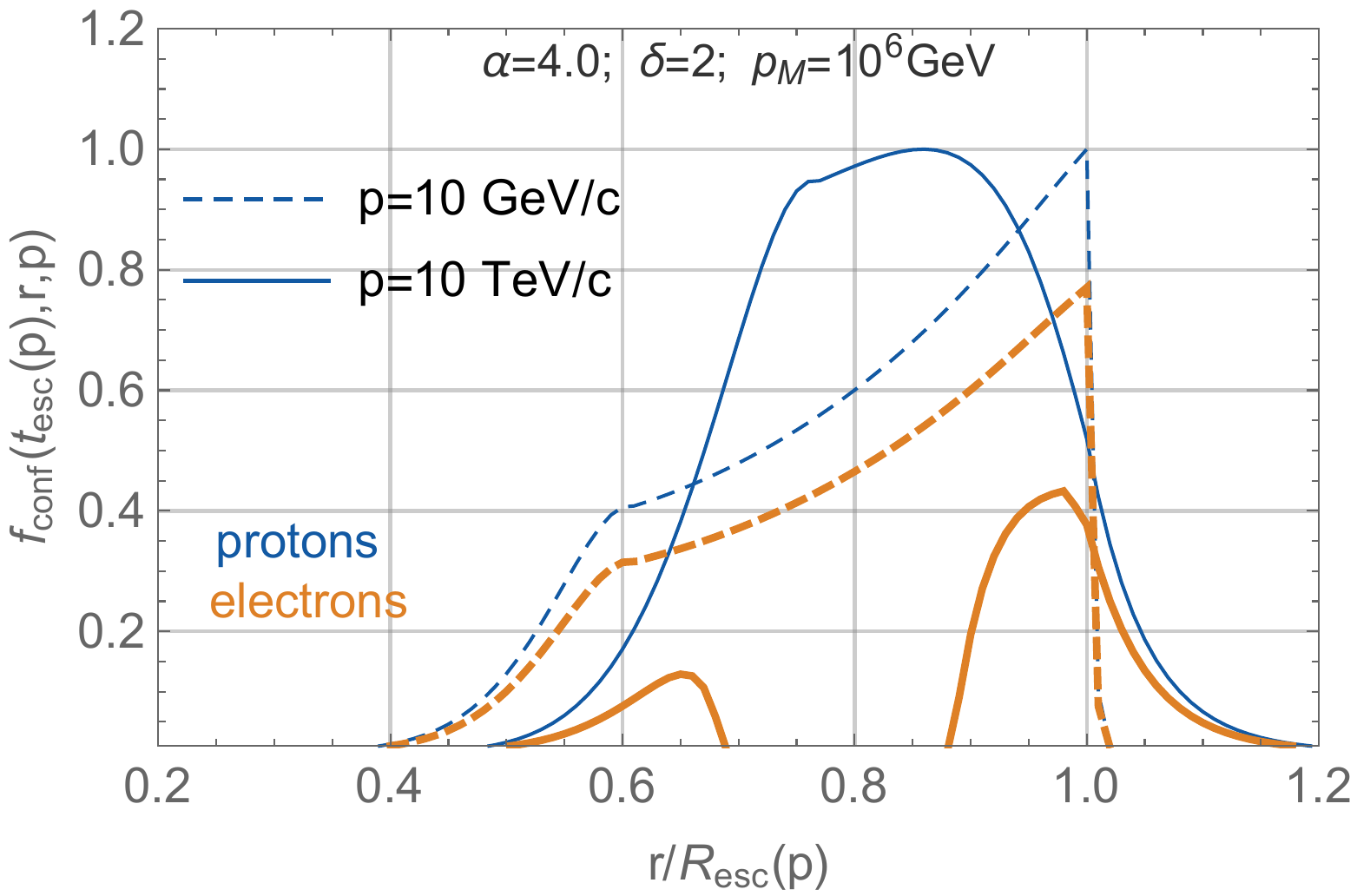}
\caption{Distribution of protons (blue-thin lines) and electrons (orange-thick lines) confined inside the SNR at the time $t_{\rm esc}(p)$, namely just before their release into the ISM. The plot shows only two energies, i.e. 10 TeV and 10 GeV, which are released at $10 t_{\rm Sed} \simeq 5$~kyr and $t_{\rm SP}\simeq 50$~kyr respectively.}
\label{fig:fconf} 
\end{figure}

Because of energy losses, electrons produced at time $t^\prime$ with energy $E^\prime$ will thus have an energy 
$E(t)$ at a later time $t$ given by \citep{Reynolds:1998}: 
\begin{equation} 
\label{eq:Ee_evol}
 \frac{E(t)}{E^\prime} = \frac{L(t^\prime,t)}{1 + A \, E^\prime \int_{t^\prime}^t L(t^\prime,\tau) \left[ B_{2,\rm tot}^2(\tau) + B_{\rm eq}^2 \right] \, {\rm d}\tau} \,,
\end{equation}
where $A= \sigma_{\rm T} c/(6\pi m_{\rm e}^2 c^4)$. The electron spectrum at time $t$ can therefore be computed by imposing number conservation, namely $f_e(E) dE dV = f_{e,0}(E^\prime) dE^\prime dV^\prime$. From Eq.~\eqref{eq:Ee_evol} we have $dE^\prime/dE= L (L-IE)^{-2}$  where $I(t^\prime,t) = A \int_{t^\prime}^{t} L(t^\prime,\tau) \left[ B_{2,\rm tot}^2(\tau) + B_{\rm eq}^2\right] d\tau$. Hence the spectrum of confined electrons is 
\begin{equation} 
\label{eq:fel(t,r)}
  f_{e,\rm conf}(E,r,t) = f_{e,0} \left( \frac{E}{L(t^\prime,t) - I E} , t^\prime \right) \frac{L}{\left(L- I E \right)^2} \frac{dV^\prime}{dV}  \,,
\end{equation}
where the ratio between differential volume elements can be written as:
\begin{equation} 
\label{eq:dV}
  \frac{dV^\prime}{dV} = L^3(t^\prime,t) \,.
\end{equation}
Eq.~\eqref{eq:fel(t,r)} can be used to also calculate the confined proton distribution by vanishing the radiation losses (i.e. $I=0$). For comparison, by setting the values of the model parameters as summarized in Table~\ref{tab:tab1}, we show in Figure~\ref{fig:fconf} the distribution of confined particles with energy of 10 TeV and 10 GeV in correspondence of their escape time, which amounts $t_{\rm esc}(10~{\rm TeV}) \simeq 10 t_{\rm Sed}$ and $t_{\rm esc}(10~{\rm GeV}) \simeq t_{\rm SP}= 95 t_{\rm Sed}$, respectively. The illustration shows that the peak of proton distribution is actually located behind the shock if $p>p_{\max}(t_{\rm SP})$, because $p_{\max}(t)$ decreases faster than the particle energy inside the SNR as due to adiabatic losses (given by Eq.~\eqref{eq:Ee_evol}). The peak of the electron distribution is, instead, closer to the shock because of the radiative losses. In addition, the contribution of particles from the precursor is always negligible at lower energies, while it starts to be relevant only when $p$ approaches $p_{\rm M}$.

%%% SUBSECTION %%%
\subsection{Distribution of escaping electrons} 
\label{sec:el_esc}
When the SNR is old enough such that energy losses at the shock become negligible, the maximum electron energy will be determined by the age of the system, such that $E_{\max,e} = E_{\max,p}$ and electrons are able to escape the system just like protons do. To calculate the distribution of escaping electrons, $f_{\rm esc}(E,r,t)$, we follow the same approach used for protons in Paper I, i.e. we assume that particles decouple from the SNR and their evolution is governed by pure diffusion. The evolution of the distribution function is hence described by 
\begin{eqnarray}    
\label{eq:fesc}
    \frac{\partial{f_{\rm esc}}}{\partial t} = \frac{1}{r^2} \frac{\partial}{\partial r}
        \left[ r^2 D \frac{\partial f_{\rm esc}}{\partial r}\right] \,,
\end{eqnarray}
where, for simplicity, the diffusion coefficient $D$ is assumed uniform and stationary. Energy losses during the escape phase can be neglected if $D$ is large enough that the typical diffusion time $t_{\rm diff}=L_{\rm diff}^2/(4 D)$ is smaller than the energy loss time, i.e. $\tau_{\rm loss} \gg t_{\rm diff}$. Considering the typical length scale for diffusion equal to the SNR radius at the escape time, i.e. $L_{\rm diff} \simeq R_{\rm esc}(E)$, the previous condition can be rewritten as:
\begin{eqnarray}    
\label{eq:Dcritical}
    D \gg \frac{R_{\rm esc}^2}{4 t_{\rm loss}} 
    \simeq 4 \times 10^{24} E_{\rm TeV}^{1-4/(5\delta)} {\langle B_{\mu G} \rangle}^2 \, {\rm cm^2 \, s^{-1}} \,,
%D \gg \frac{R_{\rm esc}^2}{4 t_{\rm loss}} \simeq 10^{24} E_{\rm TeV}^{1-4/(5\delta)} {\langle B_{\mu G} \rangle}^2 \, {\rm cm^2 \, s^{-1}} \,, %\delta=3
\end{eqnarray}
where $\langle B_{\mu G} \rangle$ is the average magnetic field in the diffusion region (which includes both the interior and exterior of the remnant).
The numerical estimate in the right hand side of the equation was obtained by using the benchmark parameters summarised in Table~\ref{tab:tab1} with $\delta = 2$. It is clear that, even assuming an average magnetic field of $10 \, \mu$G, the critical value reported in Eq.~\eqref{eq:Dcritical} is about 3 orders of magnitude smaller than the average Galactic diffusion coefficient at $\sim$TeV energies (and even smaller at lower energies). Hence, unless this condition is violated, we can safely neglect energy losses at all times $t>t_{\rm esc}(p)$. 

For the sake of completeness, we report here also the full solution of Eq.~\eqref{eq:fesc} even if this is not needed for the calculation of the final spectrum released into the Galaxy.
Such a solution can be found by taking advantage of the Laplace transforms, as shown in Paper I. The initial condition at the beginning of escape for each energy $E$ is $f_{\rm esc}(E,r,t_{\rm esc}(E)) = f_{e, \rm conf}(E,r,t_{\rm esc}(E))) \equiv f_c(E,r)$, and the solution at later times reads as
\begin{equation}    
\label{eq:fesc_sol}
    f_{\rm esc}(E,r,t) =  
        \int_0^{R_{\rm esc}(E)}   
        \left[ e^{-\left( \frac{r-r'}{R_d} \right)^2} - e^{-\left( \frac{r+r'}{R_d} \right)^2} \right] \frac{f_{c}\left(E,r' \right) r'}{\sqrt{\pi} R_d r}
         dr' \,,
\end{equation}
where $R_d = \sqrt{4 D(E) \left(t-t_{\rm esc}(E) \right)}$ is the diffusion length. For a solution including also energy losses see, e.g. \cite{Ohira_el:2012}.

%Equation~\eqref{eq:fesc_sol} is similar to the one found by \cite{Ohira+2012} but there is an important difference. \cite{Ohira+2012} assume that escaping particles (both electrons and protons) are initially located only at the shock while our initial condition is different from zero also inside the SNR. Their solution can, hence, be recovered from Equation~\eqref{eq:fesc_sol} assuming $f_c(E,r) \propto \delta(r-R_{\rm esc}(E))$. Now, while at late times and for distances larger than the SNR size the two solutions become very similar, for times comparable with the diffusion time across the SNR they instead differ significantly. In Fig.~\ref{fig:fesc_ohira} we compare the two solutions at different times after $t_{\rm esc}(E)$. Such a difference will reflect into a different shape of X-ray and $\gamma$-ray emission.

%\begin{figure}
%\centering
%\includegraphics[width=0.47\textwidth]{figures/fesc_ohira.pdf}
%\caption{Comparison between our solution for escaping particles, Eq.~\eqref{eq:fesc_sol} (solid lines), and the one used in \citet{Ohira+2012} (dashed lines). The three different colors show different times as reported in the legend.}
%\label{fig:fesc_ohira} 
%\end{figure}

%%% SECTION %%%
%%% The spectrum released into the Galaxy %%%
\section{The spectrum released into the Galaxy}
%\section{Results} 
\label{sec:results}
In this Section, we calculate the total electron and proton spectra released by a single SNR similarly to what we previously only did for protons in Paper I. 
As discussed in the previous Section, after $t_{\rm esc}(p)$ we consider particles with momentum $p$ completely decoupled from the SNR evolution, and negligible energy losses for $t > t_{\rm esc}(p)$. Under these assumptions, the total distribution of CRs with momentum $p$ injected into the Galaxy by an individual SNR is given by two contributions: particles contained inside the SNR at the time of escape, $N_{\rm inj}^{\rm dw}$, plus particles located in the shock precursor, $N_{\rm inj}^{\rm pr}$, i.e.
\begin{eqnarray}
\label{eq:finj}
    N_{\rm inj}(p) = N_{\rm inj}^{\rm dw}(p) + N_{\rm inj}^{\rm pr}(p)
    \simeq \int_0^{R_{\rm esc}(p)} 4\pi r^2 f_c(p,r) dr + \nn \\
    \,+\,
    4 \pi R_{\rm esc}^2(p) 
    \frac{D_1\left(p,t_{\rm esc}(p)\right)}{u_{\rm sh}\left(t_{\rm esc}(p)\right)} 
    f_0 \left(p,t_{\rm esc}(p)\right)\,, \hspace{1cm}
\end{eqnarray} 
where for the latter contribution we considered a precursor thickness of $D_1(p)/u_{\rm sh} \ll R_{\rm sh}$, and a diffusion coefficient in the precursor calculated as $D_1=D_B/\mathcal{F}$ through Eq.~\eqref{eq:F}.
As we already discussed in \S~\ref{sec:evolution}, we assume that the acceleration process stops when the SNR enters the radiative phase. Hence all particles having $p<p_{\rm max}(t_{\rm SP})$ are released instantaneously at $t_{\rm SP}$. Note that the contribution from the precursor is relevant only at momenta close to $p_{\rm M}$. For this reason we neglected such a contribution in the computation of the injection spectrum performed in Paper I, where we were mainly interested in calculating the slope of the injection spectrum, while we here account for it as well in order to model spectral features in the cut-off region too, and compare these to the observed electron spectrum.

In the following subsections we will discuss separately the effects induced on the electron spectrum by the self-generated magnetic field (\S~\ref{sec:results_SI}), magnetic field amplified through MHD instabilities (\S~\ref{sec:results_MHD}) and time dependent injection (\S~\ref{sec:results_Kep}). Table~\ref{tab:tab1} summarizes the values of model parameters adopted as benchmark case, while Table~\ref{tab:tab2} details about specific values adopted in the different Figures shown.
Note that for our benchmark case we have assumed the slope of acceleration spectrum $\alpha=4$, even though SNR observations favour steeper spectra with $\alpha\simeq 4.2-4.3$. However our conclusions do not change significantly for steeper acceleration spectra, as discusses in \S\ref{sec:results_all} where we used $\alpha=4.2$.
%
%\begin{table}
%\centering
%\footnotesize
%\caption{Benchmark values for the set of parameters describing the SNR evolution and the particle acceleration model.}
%\label{tab:tab1}
%\begin{tabular}{ccccccc}
%\hline	
%$E_{\rm SN}$  &  $M_{\rm ej}$  &  $n_0$  &  $\xi_{\rm CRp}$  & $\alpha$ & %$p_{\rm M}/c$ & $\delta$  \\  
%\hline
%$10^{51}$ erg  & 1 $M_{\odot}$ &  0.1 cm$^{-3}$  & $10\%$  & 4 & 1 PeV & 2 \\
%\hline
%\end{tabular}
%\end{table}
%%
%%%%%%%%%%%%%%%%%%
\begin{table}
\centering
\footnotesize
\caption{Benchmark values for the set of parameters describing the SNR evolution and the particle acceleration model.}
\label{tab:tab1}
\begin{tabular}{llc}
\hline	
Symbol  &  description  &  benchmark value \\
\hline	
$E_{\rm SN}$  &  kinetic energy of SN explosion &  $10^{51}$ erg \\
$M_{\rm ej}$  &  ejecta mass         &  $1\,M_{\odot}$ \\
$n_0$         &  external particle number density    &  0.1~cm$^{-3}$ \\
$B_0$         &  external magnetic field      &  3~$\mu$G \\
$t_{\rm Sed}$ &  beginning of Sedov phase     &  518~yr       \\
$t_{\rm SP}$  &  beginning of radiative phase &  49.5~kyr\\
$\xi_{\rm CRp}$  & proton acceleration efficiency & 0.1       \\
$\alpha$      &  slope of accelerated particles & 4   \\
$p_{\rm M}$   &  proton maximum energy at $t_{\rm Sed}$  & 1 PeV/$c$  \\
$\delta$      &  time slope of maximum energy & 2  \\  
$\xi_{\rm B}$       &  efficiency of tur. mag. field amplific.  & 0  \\
$K_{\rm ep}$  &  electron/proton ratio at the shock  & 1  \\
$q_k$         &  slope $K_{\rm ep}(u_{\rm sh})$ ($K_{\rm ep} \propto u_{\rm sh}^{-q_k}$)      & 0  \\
\hline
\end{tabular}
\end{table}
%%%%%%%%%%%%%%%%%%
%%
\begin{table}
\centering
\footnotesize
\caption{Summary of parameters' range explored in different Figures. The missing parameters are fixed as in Table~\ref{tab:tab1}.}
\label{tab:tab2}
\begin{tabular}{c|ccccc}
\hline	
& $\alpha$ & $p_\textrm{M}$ [PeV/$c$] & $\delta$ & $\xi_{\rm B}$ & $q_{k}$ \\
\hline
Fig.~\ref{fig:EmaxBdown} & 4 & 1 & 2 & $0-2\%$ & 0 \\
Fig.~\ref{fig:fconf} & 4 & 1 & 2 & 0 & 0 \\
Fig.~\ref{fig:Ninj-SG} & 4 & $0.1-1$ & $1-2-3$ & 0 & 0 \\ 
Fig.~\ref{fig:comparison_CBC} & 4 & $0.1$ & 2 & 0 & 0 \\
Fig.~\ref{fig:Ninj-MHD} & 4 & $0.1$ & 2 & $0-10\%$ & 0 \\
Fig.~\ref{fig:Ninj-Kep}  & 4 & $0.1$ & 2 & 0 & $0-1.5$ \\
Fig.~\ref{fig:Ninj-comb} & 4.2 & $0.1-1$ & $2.2-2.5$ & $2\%-5\%$ & 1 \\
\hline  
\end{tabular}
\end{table}

%%% SUBSECTION %%%
\subsection{Results for CR-amplified magnetic field} 
\label{sec:results_SI}
\begin{figure*}
\centering
\includegraphics[width=0.9\textwidth]{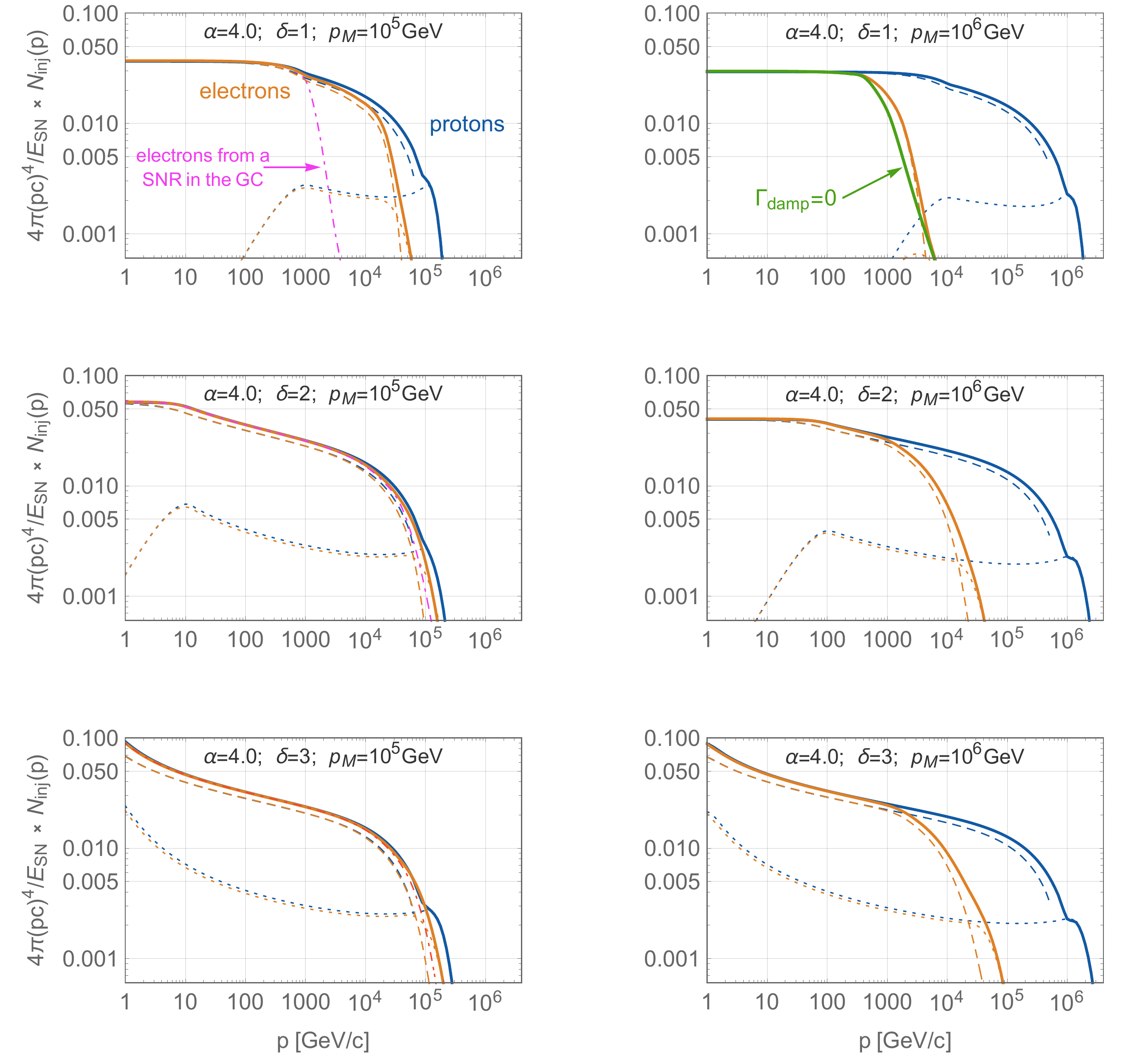}
\caption{Proton and electron spectra released by an SNR into the circumstellar medium for different values of the model parameters when the magnetic field turbulence is self-generated by CRs. All cases assume $E_{\rm SN}$, $M_{\rm ej}$, $n_0$ and $\alpha$ given by Table~\ref{tab:tab1}, while $\delta$ varies from 1 to 3 (top to bottom) and $p_{\rm M}= 10^5$~GeV (left panels) and $10^{6}$~GeV (right panels). Dashed and dotted lines represent the contributions from the SNR interior and from the precursor, respectively, while solid lines are their sum. The upper right panel shows also the electron spectrum with null magnetic field damping (green-solid line) while pink dot-dashed lines in left panel show the electron spectrum from an SNR located in the Galactic Centre.}
\label{fig:Ninj-SG} 
\end{figure*}

Here we discuss the effect of CR-generated magnetic field only.
The calculations resulting from Eq.~\eqref{eq:finj} are shown in Figure~\ref{fig:Ninj-SG}, where the electron spectrum is compared with the proton one for different values of the most relevant parameters of the model, namely $\delta$ ranging from 1 to 3, and $p_{\rm M}= 100$ TeV and 1 PeV. All other parameters are fixed as in Table~\ref{tab:tab1}.
Note that the contributions released from the SNR interior and from the precursor are shown separately with dashed and dotted lines, respectively.

Several comments are in order. First we note that both protons and electrons show a spectral break at $p_{\max}(t_{\rm ST})$ with a slope steepening above such momentum of $\sim 0.15$. As already commented in Paper I, this effect is due to the assumption that the CR acceleration efficiency, $\xi_{\rm CR,p}$, is taken constant in time. In fact, because the maximum energy decreases with time, the spectrum normalization has to increase, resulting into a larger number of particles (per unit shock surface) released at later times.

Clearly, the electron spectrum is different from the proton one only if the magnetic field is amplified strongly enough to cause relevant energy losses. Indeed, Figure~\ref{fig:Ninj-SG} shows that only when the maximum energy reaches $\sim 1$ PeV then the electron spectrum is remarkably different form the proton one. On the contrary, for $p_{\rm M} = 100$ TeV$/c$ the two spectra are almost identical with a minor difference only above $\sim 10$ TeV, particularly for the case $\delta=1$, because of the fact that the magnetic field remains amplified above $B_0$ for a longer time with respect to cases having $\delta= 2$ and 3. 

However, even in a scenario where protons reach $\sim$PeV energies, the electron spectrum injected into the Galaxy is steeper than the proton one only above $\sim 1$~TeV, hence the self-amplification seems unable to produce steeper CRe spectra down to $\sim 10$ GeV as indicated by observations.
The same conclusion holds in the absence of magnetic field damping: in fact, differences with respect to the case with damping results to be negligible. The largest variation can be appreciated for small values of $\delta$ (see green-solid line in the upper right panel of Figure~\ref{fig:Ninj-SG}), but even in this case the electron spectrum appears only marginally affected.
The spectral shape above $\sim$TeV depends on the value of $\delta$: larger values result into a less steep decrease. 

We also explored different values of $M_{\rm ej}$ and $n_0$, though our results show only slight changes with respect to these parameters. On the other hand, if we assume a much smaller value of $\delta \simeq 0$, such that the $p_{\max}$ and $\delta B_1$ remains large even at later time, the resulting CRe spectrum decreased dramatically above few tens of GeV, again at odds with observations. 

Losses due to IC scattering are negligible compared to synchrotron losses and do not affect significantly the shape of the electron spectrum in all the case analysed here. The situation may change for SNRs close to powerful stellar clusters or in the Galactic centre region, where the infrared background light is $\sim 7$ times larger than what is considered here, implying an equivalent magnetic field $B_{\rm eq} \simeq 10\, \mu$G. Such a case is shown in the left panels of Figure~\ref{fig:Ninj-SG} with pink dot-dashed lines: the effect of IC is relevant only for $\delta=1$ because electrons remains inside the remnant for longer time with respect to higher values of $\delta$.

\begin{figure}
\centering
\includegraphics[width=0.47\textwidth]{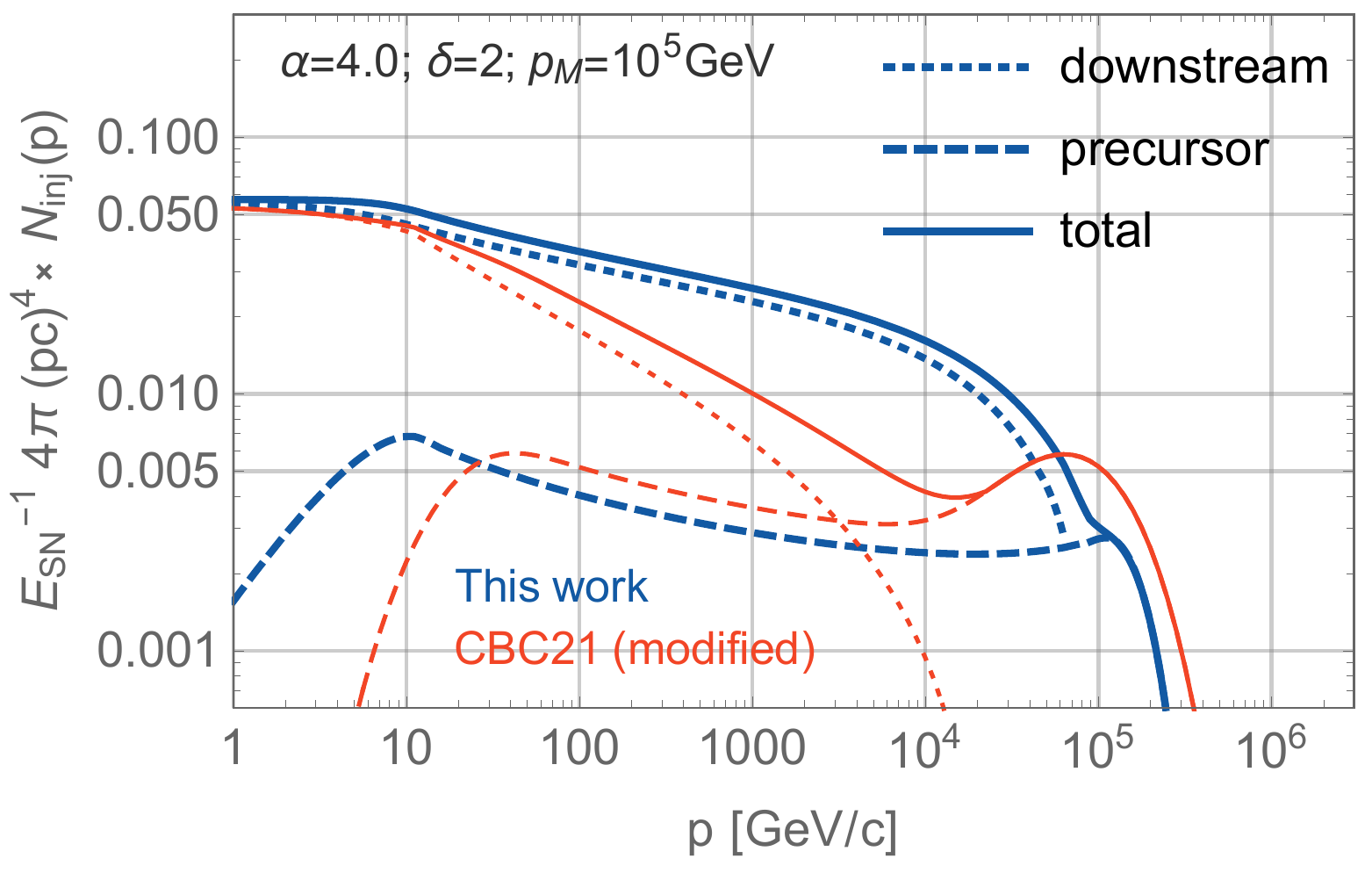}
\caption{Comparison of proton spectrum released by an SNR resulting from the computation described in this work (blue thick lines) and following the approach by CBC21 (red-thin lines), but using Eq.~\eqref{eq:pmax0} to calculate the maximum energy. Different lines show the spectrum released from the SNR interior (dotted), from the precursor (dashed), and their sum (solid). Calculations are performed assuming the benchmark case in Table~\ref{tab:tab1} with $p_{\rm M}= 100$\,TeV$/c$.}
\label{fig:comparison_CBC} 
\end{figure}

%%% COMPARISON WITH CRISTOFARI ET AL: %%%
The results of this section are compatible with the findings by \citet{Cristofari+2021} (CBC21), who show that for Type Ia SN the CRe and CRp spectra differs only for energies above $\sim$TeV. However, compared to them, the CR spectra computed here above such energy scale appear to be harder for both species. The reason for such behaviour is connected to the different method adopted for estimating $N_{\rm inj}$. Firstly, they assumed that particles advected downstream of the shock are all released at the end of ST phase, irrespective of their energy. In other words, the integral in Eq.~\eqref{eq:finj} is performed up to $R_{\rm sh}(t_{\rm SP})$ for all particle energies. As a consequence, particles suffer more adiabatic losses than in our approach. A second difference concerns the escaping flux from the precursor, which they calculated by setting a free escape boundary condition ahead of the shock (see their equation 21). Such a solution is formally appropriate only when the shock is stationary and its speed does not change with time. 
In Figure~\ref{fig:comparison_CBC} we compare the proton spectrum resulting from our computation with that resulting from CBC21 for a case with $\alpha=4$, $\delta= 2$ and $p_{\rm M}= 100$~TeV. Note that in order to show the comparison among the two approaches, the particle maximum energy is computed in both cases with Eq.~\eqref{eq:pmax0} (though the actual computation in CBC21 is performed consistently with Bell instability development). As anticipated, it is possible to see that the spectrum of particles released from the downstream appears harder in our work, because particles did suffer less adiabatic losses. The spectrum of particles escaping from the precursor is, instead, in a reasonable agreement with the exception of the bump at the highest energies, that appears more pronounced in the approach by CBC21 because they also accounted for the escape during the ED phase, while we do not.

%%% SUBSECTION %%%
\subsection{Results for turbulent amplification} 
\label{sec:results_MHD}

\begin{figure}
\centering
\includegraphics[width=0.47\textwidth]{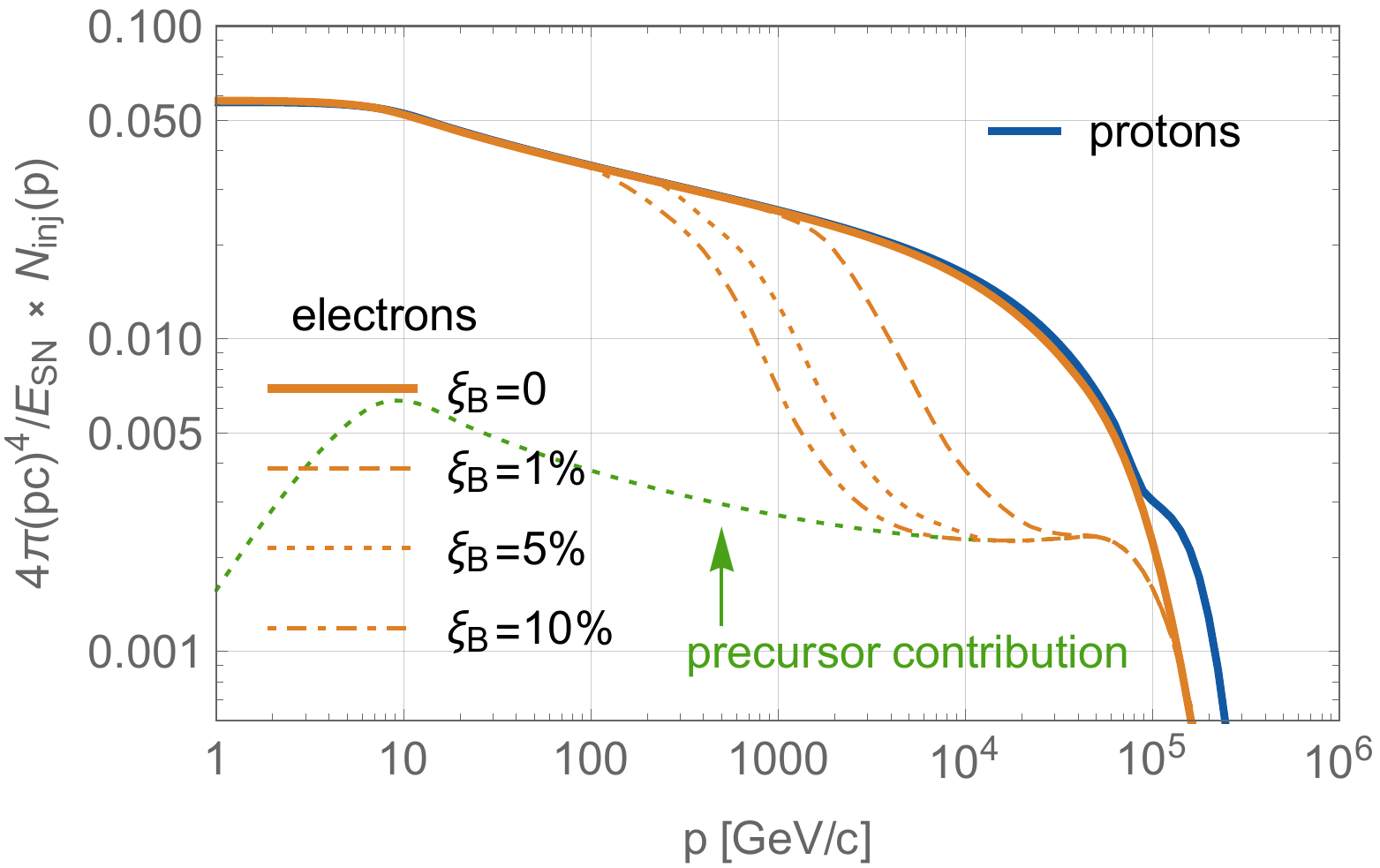}
\caption{Spectrum injected into the Galaxy for protons and electrons, accounting for the amplification of magnetic field downstream of the shock due to MHD instabilities. The green-dotted line represents the precursor contribution to electrons which is identical for all cases.  Calculations are performed assuming the benchmark case in Table~\ref{tab:tab1} with $p_{\rm M}= 100$\,TeV$/c$.}
\label{fig:Ninj-MHD} 
\end{figure}

Additional magnetic field amplified only downstream of the shock enhances the electron synchrotron losses without affecting the maximum energy reached at the shock. Following the prescription provided in \S~\ref{sec:mag-field_MHD}, we calculated electron and proton spectra released by the SNR for the same benchmark case of Table~\ref{tab:tab1}, while assuming that a fraction $\xi_{\rm B}$ of the shock kinetic energy is converted into turbulent magnetic field. In order to show the effect of turbulent amplification only, we fixed $p_{\rm M}=100$~TeV, such that losses due to CR-generated magnetic field are negligible. Results are shown in Figure~\ref{fig:Ninj-MHD}, where we explored different values of $\xi_{\rm B}$ up to 10\%. Larger values are somewhat unrealistic, in that they would imply an average downstream magnetic field strength $B_2 \gtrsim 500\,\mu$G for $t\lesssim t_{\rm ST}$, which is larger than what is estimated from observations \citep{vink2012}. 
From Figure~\ref{fig:Ninj-MHD} one can see that for $\xi_{\rm B}= 1\%$ the effect of losses is important above $\sim 1$~TeV, producing a steepening $\simeq 0.8$ up to $\sim 20$~TeV with respect to the energy range below 1 TeV. 
For $\xi_{\rm B}=10\%$, instead, losses start to be important already at $\sim 100$~GeV.
Interestingly, the spectrum flattens at higher energies due to the contribution of particles escaping from the precursor region (green-dotted line), regardless of the specific value of $\xi_{\rm B}$, because it is unaffected by losses in the SNR interior.

In conclusion, we find that turbulent amplification of magnetic field occurring downstream of the SNR shock is unable to explain the observed difference between CRe and CRp spectra down to $\sim 10$~GeV unless one assumes an unreasonably large value of $\xi_{\rm B}$. 

It is worth stressing that such a conclusion is limited to SNRs expanding through an uniform medium with density $n_{0}\lesssim 1 \, \rm cm^{-3}$. For larger densities (like those encountered at the very beginning of the evolution of a core-collapse SNR, expanding into the progenitor's wind profile) the turbulent amplification might be much more effective, possibly lowering the break energy down to $\sim 10$\,GeV. However, even in such a case the observed CRe spectrum will not be reproduced, mainly as a consequence of two different effects: firstly, because the maximum energy of escaping electrons would be reduced to values smaller that $\sim 1$\,TeV, and secondly because the slope variation at the break is $\sim 1$, much larger than the required value of 0.3. 
In principle, one can still speculate that the contributions from a population of SNRs expanding through different environments with different values of electron acceleration efficiency may combine to result into the observed CRe spectrum. While such a scenario would require a considerable level of fine tuning, it cannot be excluded. In-depth investigations are required, which we plan to perform in a future work, tailored also at exploring the evolution of magnetic fields and particle acceleration/propagation at core-collapse SNR shocks.

\subsection{Results for time dependent injection} 
\label{sec:results_Kep}
A spectral difference between electrons and protons might be obtained by invoking a different injection efficiency for the two species. On a general ground, such an assumption would be justified by the fact that the injection mechanisms allowing particles to enter the DSA are completely different for the two species. In particular, steeper electron spectra could in principle be obtained if the electron efficiency (relative to protons) were inversely proportional to the shock speed. We illustrate below a simple analytical estimate as to better explain this point.

As shown in \cite{Celli-escape1}, a quantitative estimate for the particle spectra of species {\it i} released in the Galaxy when losses are negligible, is given by
\beq
  N_{i, \rm inj}(p) \simeq \xi_{\rm CRi}\left(t_{\rm esc}(p)\right) \, u_{\rm esc}(p)^2 \, R_{\rm esc}(p)^3 \, p^{-\alpha} \, ,
\eeq
where $u_{\rm esc} = u_{\rm sh}(t_{\rm esc}(p))$.
When electron energy losses are negligible, electrons and protons of momentum $p$ escape at the same time, hence the ratio $N_{\rm e,inj}/N_{\rm p,inj}$ only depends on the injection efficiency ratio. Assuming that such a ratio is a simple power-law function of the shock velocity, we have: 
\beq \label{eq:Ne-over-Np}
  \frac{N_{\rm e,inj}}{N_{\rm p,inj}} = \frac{\xi_{\rm CRe}}{\xi_{\rm CRp}} = u_{\rm esc}(p)^{-q_k} \propto p^{-3 q_k/(5\delta)} \, ,
\eeq
where in the last equality we made the case for escape during the ST phase. Hence, the relation between the parameter $q_k$ and the slope difference between electrons and protons is 
\beq  \label{eq:qk}
  q_k = 5 \, \delta \, \Delta s_{\rm ep} /3 \,.
\eeq
Adopting the reference range for $\delta$ between 2 and 3 implies $q_k \simeq (1 \div 1.5)$ for $\Delta s_{\rm ep}\simeq 0.3$.

\begin{figure}
\centering
\includegraphics[width=0.47\textwidth]{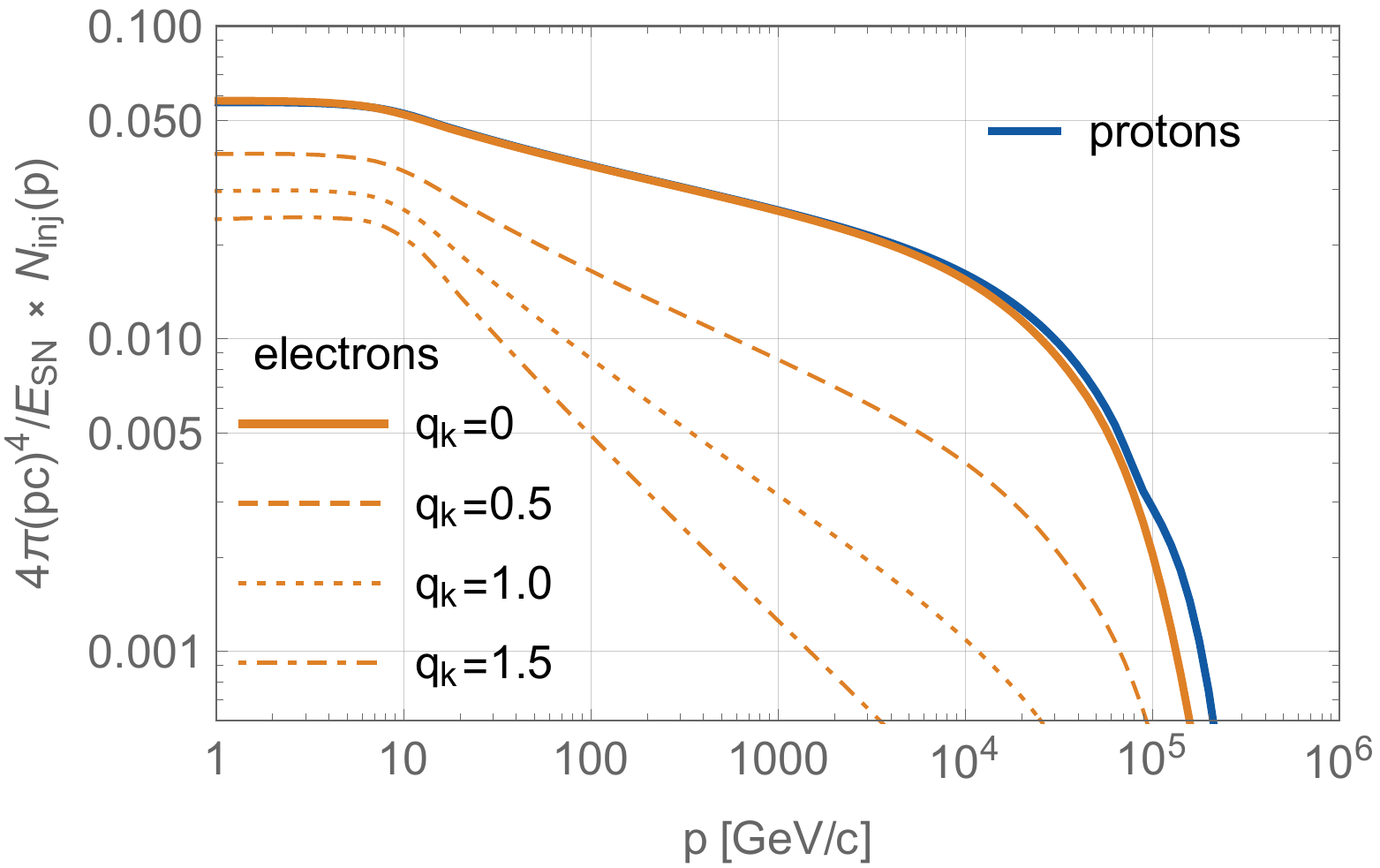}
\caption{Spectrum injected into the Galaxy for protons and electrons accounting for time dependent injection of electrons as in Eq.~\eqref{eq:Kep(t)} for different values of $q_k$ as reported in the label.}
\label{fig:Ninj-Kep} 
\end{figure}

The above analytical estimate is confirmed by the full calculation through Eq.~\eqref{eq:finj} where for the electron spectrum accelerated at the shock, Eq.~\eqref{eq:f_e0_1}, we assumed a normalization
\beq \label{eq:Kep(t)}
  K_{\rm ep} \propto {u_{\rm sh}(t)}^{-q_k} \,.
\eeq
Figure~\ref{fig:Ninj-Kep} shows the resulting $N_{\rm inj}$ for the same benchmark case of Table~\ref{tab:tab1} fixing $p_{\rm M}=100$~TeV and varying $q_k$ from 0 to 1.5. The corresponding spectral differences are in perfect agreement with Eq.~\eqref{eq:qk}, hence $\Delta s_{\rm ep}= 0.3$ is obtained for $q_k=1$. 
However, it is worth stressing that the spectral difference is only present down to $p_{\max}(t_{\rm SP})\simeq 10$~GeV, because particles at lower momenta are all released at the same time, namely the start of the radiative stage. Hence, for the purposes of reproducing the CRe spectrum observed at Earth, this mechanism is expected to be relevant only if the maximum energy at the end of the SNR life is $\lesssim 10$~GeV.

Interestingly, there are two observational evidences supporting an inverse proportionality between $K_{\rm ep}$ and $u_{\rm sh}$.
The first one is related to the multi-wavelength modelling of emission from SNRs. When enough data are available, the value of $K_{\rm ep}$ can be constrained from observations, resulting smaller ($\sim 10^{-4}\div 10^{-3}$) for young SNRs (like Tycho \citep{morlinoCaprioli}, Cas A \citep{Abeysekara+(CasA)2020}, RX J1713 \citep{Morlino+2009} and Vela Jr. \citep{Berezhko+2009}) and larger for middle-aged SNRs ($\sim 10^{-2}\div 10^{-1}$) (like W28 and W44 \citep{Zirakashvili-Ptuskin:2017}, HB21 \citep{Ambrogi+2019} and Cygnus Loop \citep{Loru+2021}). On the other hand, the electron/proton CR ratio measured at Earth is $\sim 10^{-2}$, namely in agreement with the framework where the bulk of CR electrons are accelerated in evolved SNRs.

The second argument is more involved, and it is related to the electron-to-proton temperature ratio in collision-less shock. Such a ratio can be estimated from Balmer lines emitted by SNR shocks propagating in partially neutral plasma, which suggests that $T_{\rm e}/T_{\rm p} \propto u_{\rm sh}^{-2}$ \citep{vanAdelsberg+2008}. In collision-less shocks, protons are heated by pure randomization of their bulk kinetic energy, such that $T_{\rm p} \propto m_{\rm p} u_{\rm sh}^{2}$. Hence, the inverse relationship between electron-to-proton temperature ratio at equilibrium condition and shock speed implies that the electron temperature itself is nearly constant with shock speed, and equal to $\sim 0.3$~keV  \citep{Ghavamian+2007, Rakowski+2009}. This can be explained by the fact that electrons are dynamically unimportant and can acquire energy from protons, as it also results from PIC simulations \cite[see, e.g.][]{Tran-Sironi:2020,Bohdan+2020}.
Recent hybrid simulations by \citet{Hanusch+2020} (with a population of electrons treated as test particles) also seem to confirm the decreasing trend of temperature ratio with $u_{\rm sh}$ at least up to Mach number of $\sim 20$.

Now, if the electron injection into DSA were related to the same mechanism responsible for their heating, a direct consequence would be an electron injection efficiency inversely proportional to the shock speed. 
This scenario has been investigated by \cite{Arbutina-Zekovic:2021} by means of PIC simulations. They found that both electrons and protons develop a non-thermal tail which starts at a momentum $p_{\rm inj,i} = \xi_{\rm i} p_{\rm th,i}$ where $p_{\rm th,i}$ is the thermal momentum of each species $i$. Their main finding is that $\xi_{\rm i}$ is roughly the same for both species, and that the slope of the electron spectrum above $p_{\rm inj,e}$ remains the same at all momenta, in spite of the fact that for $p_{\rm inj,e}< p <p_{\rm inj,p}$ electrons are pre-accelerated by a mechanism different than DSA.
As a consequence, \cite{Arbutina-Zekovic:2021} got the following approximate expression for the electron to proton ratio:
\beq \label{eq:Kep-Arbutina}
  K_{\rm ep} \simeq \left( \frac{m_e}{m_p} \frac{\Delta E}{\frac{3}{16} m_p u_{\rm sh}^2} \right)^{\frac{3}{2(R_{\rm sub}-1)}} 
  \propto u_{\rm sh}^{-3/(R_{\rm sub}-1)}
\eeq
where $\Delta E\approx 0.3$~keV is the energy removed from protons and added to electrons, while $R_{\rm sub}$ is the sub-shock compression ratio. Hence, for realistic values of $R_{\rm sub} \lesssim 4$,  $K_{\rm ep} \propto u_{\rm sh}^{-1}$, in remarkable agreement with Eq.~\eqref{eq:qk} which requires $q_k \simeq 1$ to explain the observed $\Delta s_{\rm ep}\simeq 0.3$.
We however remark that results from \cite{Arbutina-Zekovic:2021} deserve further investigations because their simulations are limited to high shock speed ($\sim c/3$) in a regime where thermal electrons are already relativistic, a condition far from being realized in typical SNR shocks.

%%% SUBSECTION %%%
\subsection{Combining all effects}
\label{sec:results_all}
\begin{figure}
\centering
\includegraphics[width=0.47\textwidth]{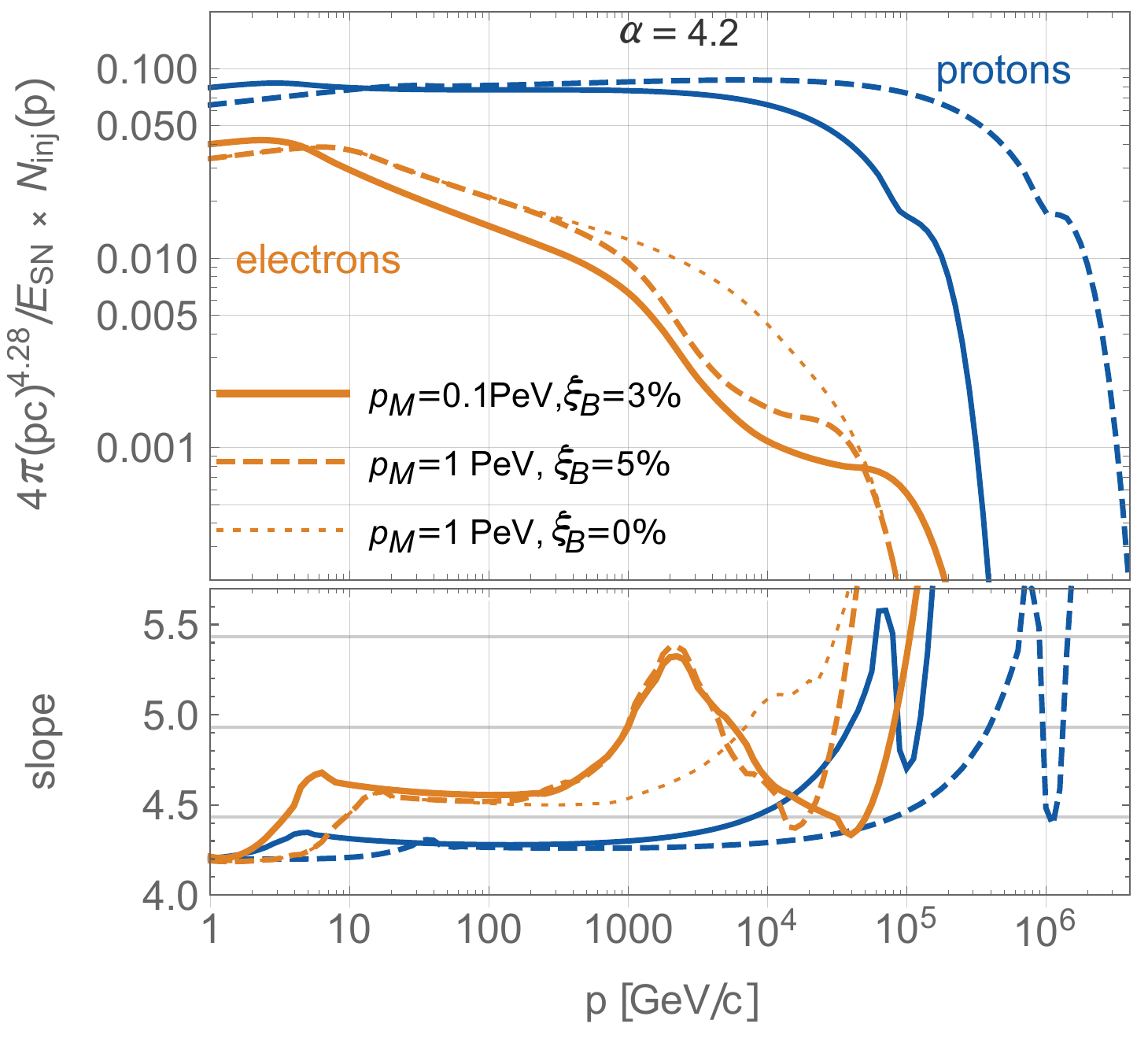}
\caption{Spectrum injected into the Galaxy of protons and electrons accounting for time dependent injection and magnetic field amplification as described in \S~\ref{sec:results_all}. Note that the spectrum is multiplied by $p^{4.28}$. The bottom panel shows the slope in momentum.}
\label{fig:Ninj-comb} 
\end{figure}

In this Section, we combine all the effects previously introduced which can impact the released spectrum, as to derive the requirements needed for reproducing the observed CRe data. This is shown in Figure~\ref{fig:Ninj-comb} which is obtained with $\alpha=4.2$, $p_{\rm M}=100$~TeV, $\delta=2.2$, $q_k=1$ and $\xi_{\rm B}=3\%$ (solid line). In such a case the CRp spectrum released into the Galaxy is $\propto p^{-4.28}$ (close to the one required once the propagation through the ISM is taken into account, see e.g. \citet{Evoli(All_CR):2019}), while the CRe is steeper by 0.3 up to 1 TeV. After this energy, losses induced by turbulent amplified magnetic field steepens the spectrum to $\propto p^{-5.4}$ up to $\sim 10$~TeV. Before the cutoff at 100~TeV, the spectrum flattens again as a result of electrons escaping from the precursor at early times. Those electrons, in fact, do not undergo adiabatic and radiative losses as the ones advected downstream.

For the case with $p_{\rm M}= 1$\,PeV, a similar result can be obtained setting $\delta= 2.5$ and $\xi_{\rm B}=5\%$, while maintaining the other parameters unchanged (dashed line). We note that the MFA resulting only from CR-instabilities is not sufficient to produce the sharp break at 1 TeV (see the dotted line representing the case of $\xi_{\rm B}=0$), such that the turbulent amplification is still required. 

%%% SECTION %%%
\section{Summary and discussion} 
\label{sec:conc}
The study of Galactic CR propagation reveals that the electron spectrum released into the Galaxy by their sources should be steeper than the proton one by $\Delta s_{\rm ep} \simeq 0.3$ in the energy range between $\sim 10$ GeV and $\sim 1$ TeV \cite[see, e.g.][]{Manconi+(electrons):2019, Evoli(electrons):2021}. At energies larger than $\sim 1$~TeV the CRe spectrum steepens further by $\sim 0.8$ \citep{HESS_CRe:2017}.
In this work we provide an interpretation for these spectral differences under the assumption that both protons and electrons are produced at SNR shocks through DSA. In such a framework,
electrons can be released with a spectrum steeper than protons by means of two different processes: either by loosing a significant amount of energy before escaping the SNR, or because the injection into DSA is not stationary, but rather characterised by a time dependent efficiency. The former case requires a large magnetic field, which might either be generated by CR-related instabilities or by turbulent instabilities developing downstream of the shock when it expands into an inhomogeneous medium.

Here we parametrically studied the aforementioned processes, as to quantify the physical requirements needed to obtain a CRe spectrum consistent with observations. In particular, for estimating the CR-generated magnetic field we assumed that the maximum momentum of protons \textbf{during the ST stage} decreases in time as $p_{\max} = p_{\rm M}(t/t_{\rm Sed})^{-\delta}$, where $p_{\rm M}$ represents the maximum momentum reached right at the Sedov time, and $\delta$ is an arbitrary parameter of the model which we constrain to be in the range $\sim 2\div 3$ because of the comparison between our model and the non-thermal emission from two middle-aged SNRs \citep{MAGIC-GammaCigni2020, Loru+2021}. 
The CR-generated magnetic field is linked to $p_{\max}$ through the condition that the acceleration time should be equal to the SNR age. Such assumption provides a level of magnetic field strength which, in turn, determines the maximum energy of electrons and the evolution of their spectrum until they are released into the Galaxy.
An additional magnetic field component that may arise from turbulent amplification is accounted for assuming that a constant fraction $\xi_{\rm B}$ of the shock kinetic energy is converted into magnetic energy. Both magnetic field components are then affected by adiabatic expansion and non linear damping.\\
In conclusion, the main results of our study are the following: \\
%\begin{enumerate}
%    \item     
(i) Losses due to magnetic field amplification (both induced by CR or turbulent instabilities) can affect the CRe spectrum only above $\sim 1$~TeV, unless the SNR expands into a very dense medium. \\
%    \item 
(ii) The CR-generated magnetic field tends to produce broad cut-offs above 1 TeV, while turbulent amplified magnetic field produces sharper spectral break, the latter being closer to the one observed in the CRe spectrum. \\
%    \item 
(iii) The spectral difference of $\sim0.3$ between CRe and CRp down to $\sim 10$~GeV can be well reproduced if the electron-over-proton injection efficiency into DSA is inversely proportional to the shock speed and, at the same time, $\delta \gtrsim 2$. \\
%    \item 
(iv) At energies $\gtrsim 10$\,TeV we predict a flattening of the CRe spectrum resulting from electrons escaped from the shock precursor, which do not suffer adiabatic and synchrotron losses inside the SNR. Such a prediction could be possibly verified with forthcoming measurements of CRe spectrum from e.g. the LHAASO observatory, which is expected to provide data extending from $\sim 500$~GeV up to 100~TeV \citep{Wu-LHAASO_el:2019}. \\
%    \item 
(v) At energies $\lesssim 10$\,GeV a spectral flattening may be present as a consequence of the stop of the acceleration at the beginning of the radiative phase. \\
%\end{enumerate}
Hence, to reproduce the entire CRe spectrum both MFA and time-dependent injection are needed, the former being more relevant at the highest energies.

Recent works have shown that the slope change above 1 TeV can be produced by a single old source located within few hundreds parsecs, like Cygnus Loop or Vela \citep{Recchia(el)2019,Fornieri+2020}. Such a scenario can only work if the Galactic diffusion coefficient is small enough such that the CRe flux at $E \gtrsim 1$\,TeV  is dominated by a single source. However, this requirement is at odds with the most recent measurements from AMS-02. In fact, as shown by \cite{Evoli(electrons):2021}, the best value of diffusion coefficient inferred from AMS-02 data is large enough that the number of sources contributing at $\sim 1$\,TeV ranges between few hundreds and a thousand. Note that this results relies on the large halo size ($\sim 5$\,kpc) estimated from Beryllium data \citep{Evoli(Beryllium):2020}.
Beyond such important caveat, it is worth mentioning that the contribution from local sources is predicted to be different whether electrons escape continuously from the source or in a burst-like event. In particular a burst-like event seems unable to reproduce the spectral break \citep{Manconi+(electrons):2019,Recchia(el)2019}. In our model, in turn, electrons are released continuously from the source, such that it is possible to find reasonable parameter values to fit the spectral break if the Galactic diffusion coefficient is tuned by hand. This exercise suggests once more that understanding particle escape from sources is of paramount importance. 

Concerning point (v), it is interesting to note that a low energy break in the energy range $\sim 1\div10$ GeV seems to be needed to not overproduce the flux detected by the Voyager 1 \citep{Cummings+2016} and the diffuse radio emission in the Galaxy \cite[see, e.g.][]{DiBernardo+2013, Orlando:2018,Vittino+2019}. 
Our findings support the existence of such a break. We note, however, that other explanations, like the spatial discreteness of CR sources \citep{Phan+2021}, may be well possible.

Finally we stress that our study is limited to SNRs expanding into uniform ISM, an assumption more suitable for type Ia SNe rather than core-collapse (CC) SNe. The reason is that the latter explode in a complex environment shaped by the prolonged activity of their progenitor's winds, which cannot be correctly treated by using purely analytical calculations.
Hence, we plan to extend this study to more complex cases in a future work. However, we believe that our main results should remain valid even for the CC SNe, essentially because the electron spectrum is mainly determined by the ST dynamical phase of the SNR, when CC and type Ia SNe do not differ substantially. The  exceptions may concern the highest ($E\gtrsim 10$\,TeV) and the lowest ($E\lesssim 10$\,GeV) energy parts of the spectrum, which are determined by the very initial and the very final stages of the SNR evolution, respectively.

\section*{Acknowledgments}
GM is grateful to Luke Drury for a stimulating discussion during the ICRC 2017 which triggered this study. The authors also acknowledge fruitful discussions with L. Sironi, E. Amato, P. Blasi, C. Evoli and P. Cristofari. We also thank S. Manconi and F. Donato for reading the manuscript and providing useful comments.
GM was supported by Grants ASI/INAF n.2017-14-H.O, SKA-CTA-INAF 2016, INAF-Mainstream 2018 and by the National Science Foundation under Grant No. NSF PHY-1748958.

%%%%%%%%%%%%%%%%%%%%%%%%%%%%%%%%%%%%%%%%%%%%%%%%%%
\section*{Data Availability}

No data have been analyzed or produced in this work. 
The phenomenological predictions performed in this work are compared with the data produced and analyzed in available publications.

\bibliographystyle{mnras}
\bibliography{References}

%%%%%%%%%%%%%%%%%%%%%%%%%%%%%%%%%%%%%%%%%%%%%%%%%%

\appendix

\section{Time evolution of shocked plasma}
\label{sec:appA}
In this Appendix, we provide further details on the implicit equations that have been set to derive the time $t^\prime(t,r)$ when the plasma element located at time $t$ in the position $r$ has been shocked. To perform this computation, we start with Eq.~\eqref{eq:ush}, rewritten in the following form
\beq
\label{eq:ode}
\frac{dr}{r}=\left(1-\frac{1}{\sigma} \right) \frac{u_{\rm sh}(t)}{R_{\rm sh}(t)} dt  \,.
\eeq
By taking advantage of Eqs.~\eqref{eq:ST_R}-\eqref{eq:ST_u} for the shock radius and speed, we now define an additional function $g(t)=u_{\rm sh}(t)/R_{\rm sh}(t)$ as
\begin{equation}
g(t) = 
\begin{dcases}
\frac{1}{t} \left[1+1.72 \left(\frac{t}{t_{\rm ch}} \right)^{3/2} \right]^{-1} \qquad t<t_{\rm Sed} \\
\frac{0.569}{t_{\rm ch}} \left[1.42 \left(\frac{t}{t_{\rm ch}} \right)-0.254 \right]^{-1} \qquad t \geq t_{\rm Sed}
\end{dcases}
\end{equation}
Integrating the LHS of Eq.~\eqref{eq:ode} between $R_{\rm sh}(t^\prime)$ and $r$ (and the RHS between $t^\prime$ and $t$), we get
\beq  \label{eq:ode_sol}
  \left(\frac{r}{R_{\rm sh}(t')}\right)^{\frac{\sigma}{\sigma-1}}
  = \frac{T_2}{T_1} \left(\frac{1+1.72\, T_1^{\frac{3}{2}}}{1+1.72\, T_2^{\frac{3}{2}}}\right)^{\frac{2}{3}}
   + \left(\frac{1.42 \,T_4-0.254}{1.42\,  T_3-0.254}\right)^{0.4}
\eeq
where $T_1 = \min[t',t_{\rm Sed}]$, $T_2 = \min[t,t_{\rm Sed}]$, $T_3 = \max[t',t_{\rm Sed}]$ and $T_4 = \max[t,t_{\rm Sed}]$.
Note that Eq.~\eqref{eq:ode_sol} is an implicit equation in $t^\prime$, that we solved by means of standard numerical techniques.

\end{document}